\documentclass{optica-article}

\journal{oe}

\articletype{Research Article}

\usepackage{lineno}
\usepackage{threeparttable} 
\usepackage{graphicx}

\begin{document}

\title{Broadband Waveguide-Coupled Photodetectors in a Submicrometer-Wavelength Silicon Photonics Platform}

\author{Alperen Govdeli,\authormark{1,2,4} Jared C. Mikkelsen,\authormark{1} Engjell Bebeti,\authormark{1} Hongyao Chua,\authormark{3} Guo-Qiang Lo,\authormark{3} Joyce K. S. Poon,\authormark{2} and Wesley D. Sacher\authormark{1,5}}

\address{\authormark{1}Max Planck Institute of Microstructure Physics, Weinberg 2, 06120 Halle, Germany \\
\authormark{2}Department of Electrical and Computer Engineering, University of Toronto, 10 King's College Road, Toronto, Ontario, M5S 3G4, Canada\\
\authormark{3}Advanced Micro Foundry Pte Ltd, 11 Science Park Road, Singapore Science Park II, 117685, Singapore\\
\authormark{4}alperen.govdeli@mpi-halle.mpg.de\\
\authormark{5}wesley.sacher@mpi-halle.mpg.de}

\begin{abstract*}
Advances in silicon (Si) photonics at submicrometer wavelengths are unlocking new opportunities to realize miniaturized, scalable optical systems for biophotonics, quantum information, imaging, spectroscopy, and displays. Addressing this array of applications with a single integrated photonics technology requires the development of high-performance active components compatible with both visible and near-infrared light. Here, we report waveguide-coupled photodetectors monolithically integrated in a foundry-fabricated, short-wavelength, Si photonics platform. We demonstrate two detector variants that collectively cover a continuous wavelength span of $\lambda =$ 400 - 955 nm. The devices exhibited external quantum efficiencies exceeding 60\% and 12\% over 400 - 748 nm and 749 - 955 nm wavelength ranges, respectively. Measured dark currents were $<$ 2 pA at a 2 V reverse bias. High-speed measurements at $\lambda =$ 785 nm demonstrated optoelectronic bandwidths up to 18 GHz. Avalanche operation was characterized, yielding a gain-bandwidth product of 374 GHz.

\end{abstract*}

\section{Introduction}

Recent progress in silicon (Si) photonics at short wavelengths ($\lambda <$ 1000 nm) is expanding the capabilities of integrated photonic microsystems, making possible a range of applications previously beyond the reach of conventional Si photonics operating at 1310/1550 nm telecommunication wavelengths. These applications span the visible (VIS) and near-infrared (NIR) spectra, and include quantum information (with trapped ions, neutral atoms, and diamond color centers) \cite{mehta2020integrated,blumenthal2024enabling,katsumi2025recent,maring2024versatile, dong2022high, palm2023modular, puckett2021422, bogaerts2020programmable, arrazola2021quantum}, medical imaging and research \cite{rank2021toward,archetti2019waveguide,diekmann2017chip,hosseini2025focusing}, biosensing \cite{rank2021toward, michalet2014silicon,fan2008sensitive,melnik2016local,nitkowski2011chip}, neurotechnology \cite{mohanty2020reconfigurable,roszko2025foundry,Mu2025Nanophotonic_CLEO}, microdisplays \cite{raval2018integrated,shi2024flat}, and underwater communication \cite{wu2017blue,notaros2023liquid}. From the outset, addressing a broad set of these applications within a single foundry-fabricated, generic Si photonics platform operating at submicrometer wavelengths presents a practical path toward technological maturation, minimizing the cost and time associated with developing individually customized platforms, while supporting scalable manufacturing. A central challenge to this approach is the development of a complete set of passive and active photonic components spanning this wavelength range (exceeding an octave). The broad optical transparency of silicon nitride (SiN) and aluminum oxide (Al$_2$O$_3$) waveguides, now readily fabricated on 200- \cite{sacher2023active,Engjell2025CLEO} and 300-mm wafers \cite{west2019low,sorace2018multi}, has enabled a growing list of short-wavelength device demonstrations: low-loss waveguides \cite{chauhan2022ultra,sacher2019visible,xiang2021high,liu2024tunable}, fiber-to-chip couplers \cite{sacher2023active}, beam combiners \cite{sacher2019visible}, multiplexers \cite{Shah2024}, polarization management devices \cite{hattori2024integrated}, optical phase shifters/switches \cite{yong2022power,notaros2022integrated,dong2022high,mohanty2020reconfigurable,liang2021robust,li2022high,dong2022piezo,govdeli2025integrated}, hybrid-integrated lasers \cite{corato2023widely,mu2025flip}, and waveguide-coupled photodetectors (PDs)  \cite{yanikgonul2021integrated,yanikgonul2022high, lin2022monolithically,morgan2021waveguide,chatterjee2019high,chatterjee2020compact}. While these demonstrations encompass the essential building blocks of integrated photonics, each has been limited to operation in either the VIS or NIR spectral range. Realizing these capabilities across both VIS and NIR wavelengths --- all within a single photonic platform --- remains an outstanding challenge.

The development of broadband PDs with high responsivities, low dark currents, and large optoelectronic bandwidths offers a versatile pathway for enabling on-chip functions critical to the aforementioned applications, including power monitoring for control of complex photonic integrated circuits (PICs), high-speed detection, and low-light-level sensing. Owing to the strong optical absorption of Si at submicron wavelengths, the Si layer(s) in integrated photonic platforms are a natural choice for the monolithic integration of short-wavelength PDs. Prior studies have reported both evanescent and end-fire coupling from low-loss SiN \cite{yanikgonul2021integrated,yanikgonul2022high,lin2022monolithically,chatterjee2019high,chatterjee2020compact} or Al$_2$O$_3$ \cite{morgan2021waveguide} waveguides to Si PDs. Although high external quantum efficiencies (EQEs) have been demonstrated, these implementations have been limited to discrete wavelengths: $\lambda=$ 685 nm \cite{yanikgonul2021integrated}, $\lambda=$ 850 nm \cite{yanikgonul2022high,chatterjee2019high,chatterjee2020compact}, and six selected VIS wavelengths \cite{lin2022monolithically}. The design and characterization of these devices for broad optical bandwidths present a key challenge for further advancement of integrated photonics for submicrometer wavelengths.

Here, we introduce SiN waveguide-coupled, silicon-on-insulator (SOI) PDs, achieving broadband operation over a continuous spectral range of $\lambda$ = 400 - 955 nm. These devices are monolithically integrated into a foundry-fabricated Si photonics platform designed for submicrometer wavelengths. Building on our prior work \cite{sacher2023active,mu2025flip}, the current iteration of our photonic platform features three SiN waveguide layers with distinct thicknesses (70, 120, and 250 nm), connected via low-loss inter-layer adiabatic transitions \cite{sacher2023active,Engjell2025CLEO}. This multilayer architecture enables the realization of photonic devices of both moderate and weak optical confinement for VIS and NIR light. PDs are realized in our platform through optical coupling from SiN waveguides to underlying Si junctions, leveraging hybrid modes spanning both layers. By employing full- and partial-etch steps in the SiN layer closest to the Si junctions, we tailor the optical coupling and demonstrate two PD variants, one optimized for blue/green wavelengths and the other for red/NIR operation. Together the PDs achieve external quantum efficiencies $>$60\% and $>$12\% over wavelength spans of 400 - 748 nm and 749 - 955 nm, respectively. Additionally, the devices exhibit low dark currents $<$ 2 pA (at a 2-V reverse bias), support optoelectronic bandwidths up to 18 GHz, and operate in avalanche mode with a gain-bandwidth product of 374 GHz. Overall, this work presents a scalable approach to broadband photodetector integration, spanning VIS and NIR submicrometer wavelengths, within foundry-fabricated Si photonics. The demonstrated device performance and diverse operating modes highlight the potential of this technology to enable short-wavelength PICs that leverage arrays of PDs to realize complex optoelectronic functionalities. %
\section{Photodetector Design}

\begin{figure} [!t] 
    \centering
     \includegraphics[width=\textwidth]{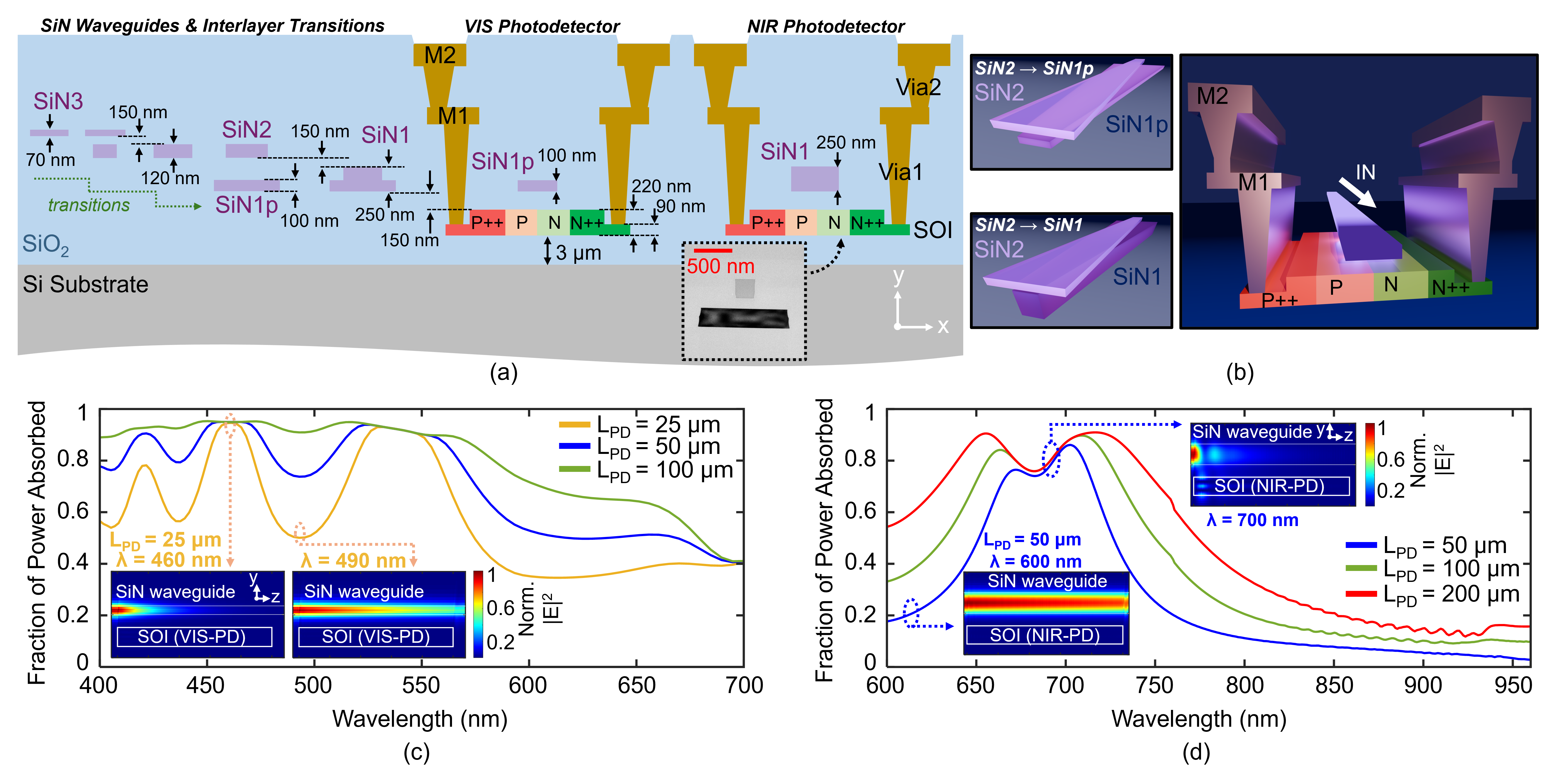}
    \caption{SiN-on-SOI waveguide-coupled PDs for visible (VIS) and near-infrared (NIR) light. (a) Photonic platform cross-section showing the SiN waveguide layers (SiN1, SiN1p, SiN2 and SiN3) and the two (VIS and NIR) photodetector variants. VIS-PDs and NIR-PDs utilize SiN1p (i.e., partially-etched SiN1) and full-thickness SiN1, respectively. Inset: Transmission electron micrograph of SiN1 atop the SOI layer. (b) Schematic perspective views of interlayer transitions between SiN waveguide layers (left) and a PD (right), not to scale. (c,d) Simulated fraction of input light absorbed within the PD junction (absorbed power ratio) vs. wavelength ($\lambda$) for (c) VIS-PDs and (d) NIR-PDs; TE-polarization, PIN-junction configuration. Traces for various PD lengths, $L_{PD}$, are shown. Insets: Simulated optical propagation within the PDs (axial cut) at various wavelengths. (a) was adapted from our conference abstract (Ref. \citenum{govdeli2025CLEO}).}
    \label{fig:Schematic_PD}
\end{figure}

Figure \ref{fig:Schematic_PD}(a) illustrates the cross-sectional schematic of our submicrometer-wavelength Si photonics platform, including three SiN waveguide layers (SiN1, SiN2, SiN3). The platform features two PD variants, VIS- and NIR-PDs, each implemented with a waveguide in the bottom SiN layer (SiN1), positioned above and in close proximity to a patterned Si detector region. The Si region consists of a 220-nm-thick, 16-\textmu m-wide central rib and a 90-nm-thick partially etched slab extending 8 \textmu m on either side. PN- and PIN-junction PDs utilize heavily-doped (P++ and N++) Si regions to form ohmic contacts, separated by a 2-\textmu m gap. For PN-junction devices, additional moderately-doped (P and N) sections are included to define the junction. The simulated PN-junction doping profile is shown in Fig. \ref{fig:TCADresults} (see Appendix). Electrical connections are established through two metal layers (M1 and M2) and corresponding via layers (Via1 and Via2). Via1 contacts the PDs on the partially-etched regions of the Si. Optical connections to fiber-to-chip edge couplers at the chip facet are implemented using routing waveguides with bends that orient the PDs orthogonally to the facet [Fig. \ref{fig:Measurement_DynamicRange}(a)], which mitigates stray input light incident on the active area of the devices. The PD test structure routing waveguides were formed from their respective SiN layers: SiN1p for VIS-PDs and SiN1 for NIR-PDs. Although not used here, adiabatic interlayer transitions in the platform can couple light between waveguide layers, Fig. \ref{fig:Schematic_PD}(b), with measured losses per transition of 0.8–2.0 dB ($\lambda = 445$–650 nm, SiN1p–SiN2), 0.3–0.6 dB (445–650 nm, SiN2–SiN3), $<0.3$ dB (775–980 nm, SiN1–SiN1p), and $<0.15$ dB (775–980 nm, SiN1–SiN2) \cite{Engjell2025CLEO}. 

The operation of the PDs is based on optical coupling between the SiN waveguide and the underlying Si absorption region, Fig. \ref{fig:Schematic_PD}(b). When light propagating in the input SiN waveguide reaches the PD, it couples into the supermodes of the SiN-on-Si structure. As these supermodes propagate, a portion of the optical field extends into the Si, where it is absorbed, generating electron-hole pairs and producing a photocurrent ($I_{ph}$). The efficiency of this process depends on the modal overlap between the guided modes and the detector’s active region (PN or PIN junction), as well as on the wavelength-dependent absorption coefficient of Si. To optimize the responsivity across the VIS and NIR spectral regions, two PD variants were implemented with waveguide geometries tailored for their respective wavelength bands. To enable efficient coupling and absorption of blue/green light, the VIS-PDs use 500-nm-wide ($W_{WG}$), 100-nm-thick SiN waveguides defined by partial etching of the SiN1 layer (SiN1p). In contrast, the NIR-PDs employ full-thickness (250-nm) SiN1 waveguides with an 800-nm width, where the enlarged waveguide dimensions are designed for coupling and absorption at longer red and NIR wavelengths.

In the remainder of this section, the expected device performance is examined through light propagation simulations, while the operating principles are illustrated using optical mode simulations. The expected absorption of the PDs was evaluated via three-dimensional finite-difference time-domain (3D-FDTD) simulations (PIN-junction configuration) for transverse-electric (TE) polarized light. For each type of PD and at various junction lengths ($L_{PD}$), we computed the fraction of optical power absorbed in the 2-\textmu m-wide intrinsic region of the junctions over the target spectral ranges, Figs. \ref{fig:Schematic_PD}(c) and \ref{fig:Schematic_PD}(d). For VIS-PDs, increasing the junction length enhances absorption, particularly in the blue/green spectral range ($\lambda$ = 400–570 nm), where nearly complete absorption $>$88\% is achieved for $L_{PD} \geq 100$ \textmu m, Fig. \ref{fig:Schematic_PD}(c). The absorption decreases at longer wavelengths, reaching $\approx$40\% at $\lambda =$ 700 nm, owing to the reduced absorption coefficient of Si and the supermode characteristics associated with the thin and narrow SiN dimensions of the VIS-PD. Importantly, localized absorption peaks are present at specific wavelengths and are particularly pronounced for short devices ($L_{PD} = 25$ \textmu m). For NIR-PDs, the larger SiN waveguide dimensions enhance absorption at red and NIR wavelengths, Fig. \ref{fig:Schematic_PD}(d). A distinct absorption peak is observed in the shortest device ($L_{PD} = 50$ \textmu m), increasing absorption around $\lambda =$ 650–740 nm. With larger PD length, this peak evolves into a broader feature with two maxima separated by a shallow dip. For $L_{PD} = 200$ \textmu m, this spectral feature yields absorption exceeding 55\% over $\lambda =$ 600-768 nm, gradually decreasing to 16\% at 960 nm. Generally, longer $L_{PD}$ enhances absorption across $\lambda =$ 600–960 nm.

\begin{figure} [!t] 
    \centering
     \includegraphics[width=0.79\textwidth]{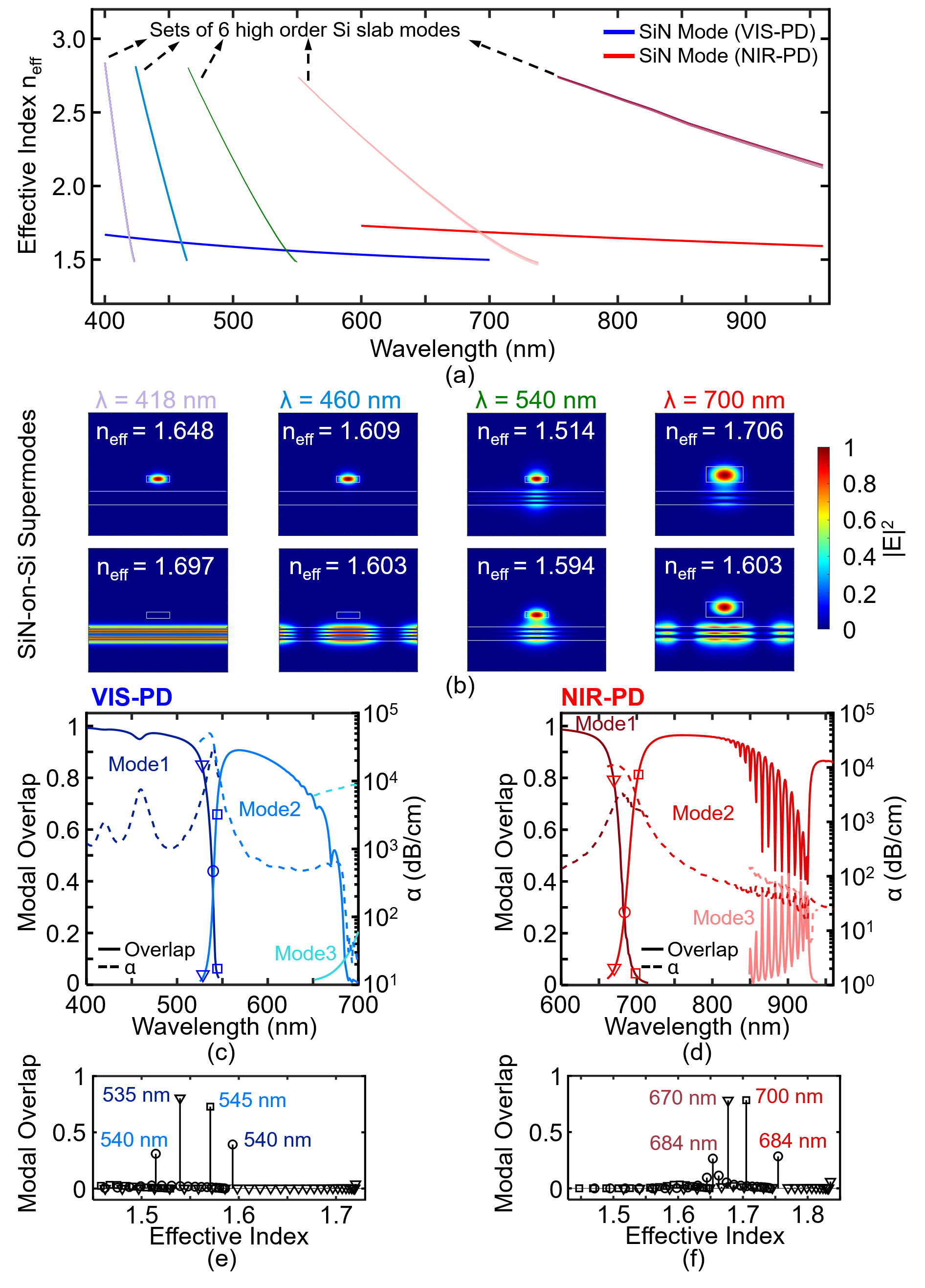}
    \caption{Operating principle of the PDs as illustrated by optical mode simulations (TE polarization, PIN PD). (a) Effective index vs. $\lambda$ crossing plot of optical modes in the SiN waveguide and Si slab; each in isolation. The fundamental SiN modes are shown for the VIS- (blue) and NIR- (red) PDs; sets of 6 high-order modes of the Si slab exhibiting index crossings with the SiN modes are also shown. (b) Mode intensity profiles ($|E|^2$) of the two supermodes of the SiN-on-Si PD exhibiting the highest overlap with the incident mode of the input SiN waveguide. The wavelengths correspond to the index crossings of (a); the first 3 belonging to the VIS-PD and the last for the NIR-PD. (c,d) Modal overlap with the input SiN mode (solid traces) and absorption loss $\alpha$ (dashed traces) vs. $\lambda$ for supermodes of the (c) VIS- and (d) NIR-PD; the three modes of highest overlap are shown; ordered from highest (Mode 1) to lowest (Mode 3) effective index. (e,f) Modal overlap with the input mode vs. effective index at various wavelengths at and near the index crossings of (a) for (e) VIS- and (f) NIR-PDs; the markers delineate the mode number and correspond to those of (c,d).} 
    \label{fig:ModeSimulations_VISPD_NIRPD}
\end{figure}

The mechanisms governing the magnitude and shape of the PD absorption spectra, and their dependence on the SiN waveguide dimensions, are illustrated through finite-difference eigenmode (FDE) simulations (TE polarization), Fig. \ref{fig:ModeSimulations_VISPD_NIRPD}. By calculating the optical modes of the SiN waveguide and Si detector separately [Fig. \ref{fig:ModeSimulations_VISPD_NIRPD}(a)], effective index crossings are identified between the fundamental SiN waveguide mode and sets of high-order slab modes in Si. These crossings, occurring near $\lambda =$ 418, 460, and 540 nm for the VIS-PD and around 700 nm for the NIR-PD, coincide with the absorption peaks in Figs. \ref{fig:Schematic_PD}(c) and \ref{fig:Schematic_PD}(d), indicating modal hybridization in the composite SiN–Si structure. Figure \ref{fig:ModeSimulations_VISPD_NIRPD}(b) presents the mode profiles of the two supermodes with the highest overlap with the input mode at each of these wavelengths. Increasing modal hybridization (overlap with both layers) occurs at longer wavelengths. Mode profiles reveal that supermodes at each index crossing differ in their number of vertical lobes within the Si. For a fixed Si thickness, the SiN waveguide dimensions set the index-crossing (hybridization) wavelengths. As discussed next, these dimensions also define the coupling between the PD input and the supermodes, as well as the supermodes’ overlap with the PD active region.

Given the large width of the Si detector, the SiN-Si structure supports many supermodes, and the overlap with the input waveguide mode feeding the PD defines which supermodes are excited. The modal absorption of these supermodes (within the Si junction) determines the absorption spectrum of the PD. These characteristics are shown in Figs. \ref{fig:ModeSimulations_VISPD_NIRPD}(c) and \ref{fig:ModeSimulations_VISPD_NIRPD}(d) for both VIS- and NIR-PDs, plotting the input overlap of supermodes and their absorption per unit length versus wavelength. The three supermodes with the highest overlap are shown, ordered by effective index from highest (Mode 1) to lowest (Mode 3). For the VIS-PD, at wavelengths between 400 and $\approx$535 nm, Mode 1 is the primary mode, exhibiting near-unity input overlap in addition to high absorption exceeding 1000 dB/cm; no other modes with significant overlap are present. Enhanced attenuation near $\lambda =$ 418 nm and 460 nm corresponds to the effective index crossings in Fig. \ref{fig:ModeSimulations_VISPD_NIRPD}(a), indicating increased overlap with Si due to modal hybridization. Similar behavior is observed for $\lambda =$ 545–600 nm, where Mode 2 exhibits high overlap and absorption but without distinct attenuation peaks, consistent with the absence of effective index crossings. Near $\lambda =$ 540 nm, coupling is roughly equally distributed between Modes 1 and 2, with absorption surpassing that of the shorter wavelength peaks, reaching $>10^4$ dB/cm. Across the three absorption peaks of the VIS-PD, increasing modal hybridization (and overlap with Si) with wavelength compensates for the decreasing material absorption, thereby defining the peak magnitudes in Fig. \ref{fig:Schematic_PD}(c). The flattened absorption spectrum at $L_{PD} = 100$ \textmu m [Fig. \ref{fig:Schematic_PD}(c)] results from the high baseline absorption and the strong input overlap of both Mode 1 and Mode 2, with the baseline absorption supported by the broad peaks persisting even away from central hybridization wavelengths. Representative cases near $\lambda = 540$ nm [Fig. \ref{fig:ModeSimulations_VISPD_NIRPD}(e)] show that, among the many supermodes in the SiN–Si structure, only one or two exhibit significant overlap with the PD input mode.

As shown in Fig. \ref{fig:ModeSimulations_VISPD_NIRPD}(e), the NIR-PD exhibits similar trends, augmented by the lower Si material absorption over $\lambda = $600–960 nm and the presence of only one hybridization wavelength ($\lambda \approx 700$ nm). These characteristics underlie the larger $L_{PD}$ required for near-unity peak absorption in Fig. \ref{fig:Schematic_PD}(d), as well as the shape of the absorption spectrum. Because Si absorption around $\lambda \approx 700$ nm is reduced compared with the blue–green wavelengths relevant to the VIS-PD, beating between Modes 1 and 2 is evident in the simulated propagation profile of the NIR-PD [Fig. \ref{fig:Schematic_PD}(d), inset], but otherwise strongly suppressed in the VIS-PD [Fig. \ref{fig:Schematic_PD}(c), inset].  

These results confirm that the chosen SiN thicknesses --- 100 nm for VIS-PDs and 250 nm for NIR-PDs --- are well matched to their respective target spectral ranges. Additional simulations, in which the NIR-PD geometry (250-nm-thick, 800-nm-wide SiN waveguide) is applied to the visible spectrum, show substantially lower absorption compared to the VIS-PD geometry (100-nm-thick, 500-nm-wide waveguide), even when narrowing the NIR-PD waveguide to 300 nm to reduce confinement in the SiN, Fig. \ref{fig:FDTD_comparison_WaveguideGeometry}(a) (see Appendix). This trend is consistent with the generally high visible-light confinement of the NIR-PD waveguide. Additional optical mode simulations [similar to Fig. \ref{fig:ModeSimulations_VISPD_NIRPD}(c)] further support this observation, showing that at a wavelength outside the hybridization peaks ($\lambda = 488$ nm), the relevant NIR-PD supermodes exhibit attenuation nearly an order of magnitude lower than those of the VIS-PD. Conversely, the VIS-PD geometry achieves near-complete absorption for a 100-\textmu m length across much of the visible spectrum, confirming its effectiveness for blue/green light. In further simulations (Fig. \ref{fig:FDTD_comparison_WaveguideGeometry}(b), see Appendix), the VIS-PD was evaluated across the red and NIR spectral ranges and compared with the NIR-PD. The VIS-PD exhibited lower absorption with reduced hybridization peaks, and, importantly, absorption remained below 3\% for wavelengths beyond 880 nm (compared with 16\% for the NIR-PD). This behavior persisted even when the VIS-PD SiN waveguide width was increased to 800 nm. Optical mode simulations further show that beyond $\lambda =$ 850 nm the effective index of the SiN input waveguide decreases to 1.48, approaching the SiO$_2$ cladding index, which indicates weak confinement and poor matching to the higher-confinement Si slab modes. Collectively, these results demonstrate that the selected 100-nm and 250-nm SiN thicknesses are important for achieving broadband absorption in the blue/green and red/NIR regions, respectively, by ensuring strong supermode overlap with both the Si absorption region and the input waveguide mode.

\section{Experimental Results}

\subsection{Fabrication and measurement setup}
\label{section_fabrication}

Our submicrometer-wavelength Si photonics platform was fabricated on 200-mm diameter silicon-on-insulator (SOI) wafers at Advanced Micro Foundry (AMF). The Si layer was first patterned and doped to form the photodetector regions. Both PN- and PIN-junction PDs were fabricated using heavily doped (P++ and N++) regions to form ohmic contacts. In the case of PN-junctions, additional moderately-doped (P and N) regions were introduced to define the junction. As with previous generations of our photonic platform, the SiN waveguide layers were fabricated using plasma-enhanced chemical vapor deposition (PECVD), deep ultraviolet lithography, and reactive ion etching \cite{lin2022monolithically}. Metal routing layers and corresponding vias were formed; chemical mechanical polishing (CMP) was used for layer planarization. Deep reactive ion etching was performed to define facets for fiber-to-chip coupling, after which the wafers were diced.

For current-voltage (I-V) and responsivity characterization, wavelength-tunable light from a supercontinuum laser source (NKT Photonics SuperK Fianium) connected to a tunable optical filter (NKT Photonics LLTF, $\approx$1.5-nm full width at half maximum linewidth) was coupled into (out) of the chip through cleaved polarization-maintaining (single-mode) optical fibers (Coherent PM-S405-XP and S405-XP). For broadband responsivity and external quantum efficiency (EQE) measurements, the input wavelength was swept continuously from 400 - 955 nm. On-chip optical power input to the PDs ($P_{in,PD}$) was determined from transmission measurements of reference waveguides with nominally-identical edge couplers to the PD test structures, allowing de-embedding of fiber-to-chip coupling losses. I-V traces were acquired using a sourcemeter (Keysight B2912A Precision) connected to the chips through tungsten needle probes.

The optoelectronic frequency response of the PDs was characterized with input light from a laser diode (Thorlabs S4FC785) at a fixed wavelength of 785 nm. In these measurements, the optical input was modulated using a lithium niobate electro-optic modulator (EOM, Thorlabs LNX7840A), driven by a 26.5 GHz vector network analyzer (VNA, Keysight P5025A). Electrical signals from the PDs were routed to the VNA via RF probes (GGB 40A-SG-100-EDP-N) and RF cables; a bias-tee (Pasternack PE1BT1002) was used for applying the reverse bias. The measured S21 responses were corrected by de-embedding the frequency-dependent contributions of the EOM and RF cables, followed by smoothing using a Savitzky-Golay filter \cite{orfanidis1995introduction,schafer2011savitzky}. The optoelectronic 3-dB bandwidth of the PDs was defined relative to the response measured at 1 GHz. Avalanche gain measurements of PN-junction PDs were conducted following a device stabilization procedure, in which a continuous reverse bias of 14 V was applied for $\approx$2 min prior to avalanche-mode characterization (performed with a maximum reverse bias of 13.5 V). As shown in Fig. \ref{fig:IV_drifting} (see Appendix), the stabilization procedure reduced the drift observed in the I-V characteristics of the PDs near their breakdown voltage. All measurements reported in this work were conducted at room temperature.

\subsection{Linear-mode characterization}
\label{LinearMode_EQEcharacterization}
\begin{figure} [!t] 
    \centering
     \includegraphics[width=\textwidth]{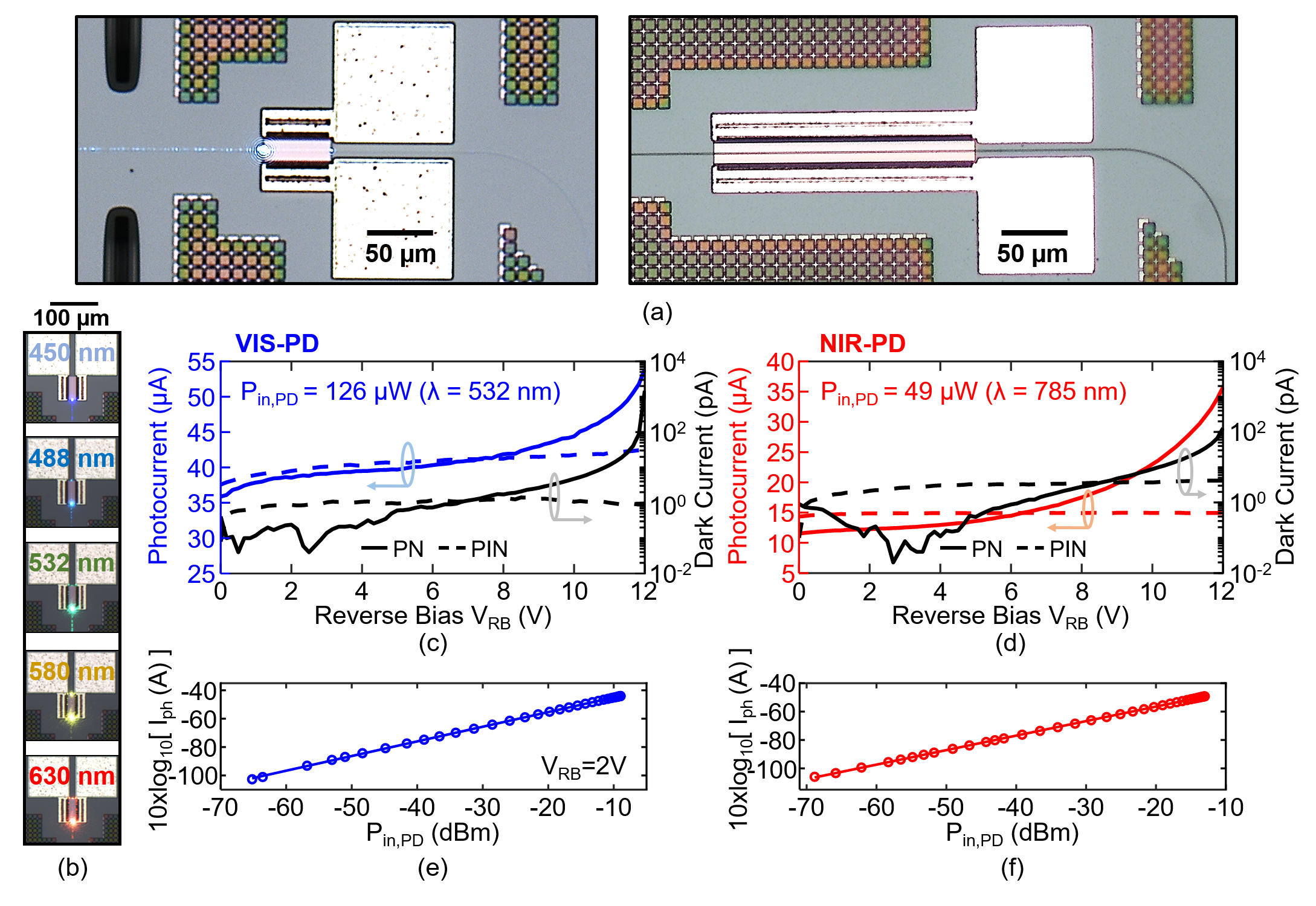}
    \caption{Current-voltage (I-V) and responsivity characterization of the PDs. (a) Optical micrographs of a fabricated VIS-PD (left) and NIR-PD (right). For improved visibility of the waveguides, the brightness and contrast have been adjusted and light was input to the VIS-PD. (b) Micrographs of a VIS-PD with various visible-wavelength optical inputs. (c,d) I-V characteristics of (c) 50-\textmu m-long VIS-PDs  and (d) 200-\textmu m-long NIR-PDs with and without input light; TE-polarization, PN and PIN configurations shown. (e,f) Photocurrent $I_{ph}$ vs. input optical power $P_{in,PD}$ for PN PDs in (c,d); reverse bias ($V_{RB}$) = 2 V. Measured data: circles, solid lines: linear fits, coefficient of determination, $R^2$, = 0.99. (b) was adapted from our conference abstract (Ref. \citenum{govdeli2025CLEO}).}
    \label{fig:Measurement_DynamicRange}
\end{figure}

\begin{figure} [!t] 
    \centering
     \includegraphics[width=0.9\textwidth]{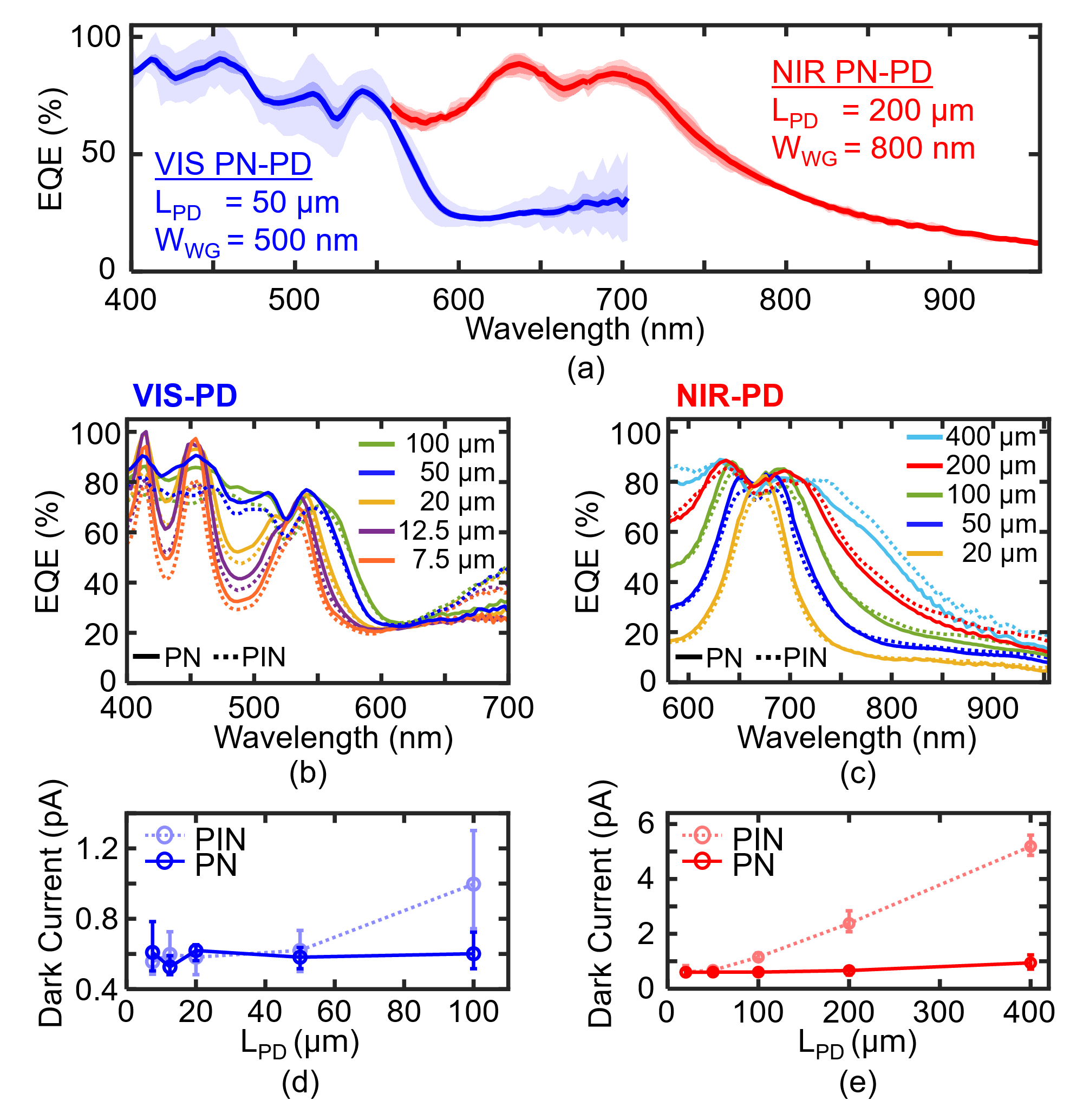}
    \caption{EQE and dark current characterization of the PDs. (a) Measured EQE spectra with TE-polarized optical inputs for VIS-PDs (6 chips) and NIR-PDs (7 chips). Solid lines: average EQE across chips, inner shaded regions: standard error, outer shaded regions: range (maximum / minimum) of the measured values. (b,c) Average EQE spectra of (b) VIS-PDs and (c) NIR-PDs vs. wavelength for different PD lengths ($L_{PD}$); PN-junction and PIN-junction devices are shown as solid and dashed lines, respectively. (d,e) Dark current vs. $L_{PD}$ for (d) VIS-PDs and (e) NIR-PDs measured across 7 chips with reverse biases of 2 V. Data points represent the mean dark current, while the bars indicate the full range (maximum/minimum) observed across all chips. Solid and dashed lines correspond to PN and PIN junction devices, respectively. (a) was adapted from our conference abstract (Ref. \citenum{govdeli2025CLEO}).}
    \label{fig:Measurement_EQEs}
\end{figure}

The PDs were first characterized in \textit{linear mode}, corresponding to operation at low reverse bias voltages ($V_{RB}$), well below their breakdown voltage ($V_{B}$) --- approximately 13.5 V for PN-junction devices and $>$20 V for PIN-junction devices. Figures \ref{fig:Measurement_DynamicRange} and \ref{fig:Measurement_EQEs} present the I-V and responsivity characteristics of the PDs. Unless otherwise specified, all measurements were performed using TE-polarized light. TM-polarization results are provided in the Appendix (Fig. \ref{fig:Measurement_EQEs_TM}). The photocurrent ($I_{ph}$) was obtained by subtracting the dark current ($I_{dark}$) from the total current measured with input light (i.e., $I_{ph}=I_{total}-I_{dark}$).

Figures \ref{fig:Measurement_DynamicRange}(c) and \ref{fig:Measurement_DynamicRange}(d) show the I-V characteristics of representative 50-\textmu m-long VIS-PDs and 200-\textmu m-long NIR-PDs, with and without optical input [$\lambda =$ 532 nm (VIS) and 785 nm (NIR)]. Responsivities, defined as $R_p=I_{ph}/P_{in,PD}$, were calculated to be 0.306 A/W for PN-type and 0.313 A/W for PIN-type VIS-PDs at a reverse bias of 2 V. Separately, the PN- and PIN-type NIR-PDs exhibited responsivities of 0.25 A/W and 0.30 A/W, respectively. For reverse bias $V_{RB}<$ 12 V, dark currents remained below 1.5 nA for the VIS-PDs and below 150 pA for NIR PDs. Also, Figs. \ref{fig:Measurement_DynamicRange}(e) and \ref{fig:Measurement_DynamicRange}(f) show the dependence of photocurrent on input optical power. For both VIS- and NIR-PDs (PN-type), the linear dynamic range exceeded 50 dB, with linear fits yielding a coefficient of determination ($R^2$) of 0.99.

Photocurrent measurements were performed across multiple samples (6 chips for VIS-PDs, 7 chips for NIR-PDs), and external quantum efficieny (EQE) measurements are presented in Fig. \ref{fig:Measurement_EQEs}. VIS-PDs ($L_{PD}$ = 50 \textmu m) exhibited EQEs exceeding 61$\pm10\%$ (mean $\pm$ standard deviation) over a $\lambda$ = 400-562 nm wavelength range, Fig. \ref{fig:Measurement_EQEs}(a), indicating efficient absorption and carrier collection in the blue/green spectral region, consistent with the simulated absorption spectra of Fig. \ref{fig:Schematic_PD}(c). The EQE progressively declined at longer wavelengths, reaching $\approx$20$\%$ at $\lambda$ = 700 nm. The NIR-PDs ($L_{PD}$ = 200 \textmu m) exhibited EQEs $>$ 60$\pm6\%$ for $\lambda$ = 560-742 nm, with decreasing EQEs at longer wavelengths ($\approx$12$\%$ at $\lambda$ = 955 nm). These results highlight the tailored spectral response of the PDs enabled by engineering of the SiN waveguide dimensions. PIN-junction devices yielded comparable spectral responses, with EQE variations within $\pm$5$\%$ compared to their PN-junction counterparts.

To evaluate the impact of PD length on device performance, EQE spectra were measured across varying device lengths, Figs. \ref{fig:Measurement_EQEs}(b) and \ref{fig:Measurement_EQEs}(c). Generally, the EQE improved with increasing $L_{PD}$, reflecting the enhanced optical absorption provided by longer PDs. Additionally, shorter VIS-PDs ($L_{PD} =$ 7.5 - 20 \textmu m) exhibited exceptionally high EQE peaks at $\lambda \approx 415$ nm, 454 nm and 540 nm, comparable to the EQE of longer devices. These EQE peaks, aligning closely with simulation results [Fig. \ref{fig:Schematic_PD}(c)], are attributed to modal hybridization with Si slab modes. However, at wavelengths outside hybridization regions, the EQE of shorter PDs declined significantly; for example, at  $\lambda$ = 487 nm, the EQE was 32\% for $L_{PD} = 7.5$ \textmu m, compared to $>70\%$ for devices with $L_{PD}>50$ \textmu m. Similar trends were observed for NIR-PDs, where modal hybridization enabled consistently high EQEs (75-85$\%$) across all tested PDs over a wavelength range of $\lambda$ = 650-700 nm. Outside this spectral window, the EQE exhibited a strong dependence on $L_{PD}$; for instance, at $\lambda$ = 600 nm, EQE increased from $\approx20\%$ for the 20-\textmu m-long PD to $\approx80\%$ for the 400-\textmu m-long device. For wavelengths beyond 715 nm, the EQE decreased across all devices, likely due to reduced Si absorption. However, PD length remained a significant factor at intermediate NIR wavelengths; at $\lambda$ = 785 nm, NIR-PDs reached EQEs of 40$\%$ and 56$\%$ for $L_{PD}$ of 200 and 400 \textmu m, respectively. 

Dark current measurements, performed at $V_{RB} = 2$ V, provided further insight into the device performance, Figs. \ref{fig:Measurement_EQEs}(d) and \ref{fig:Measurement_EQEs}(e). VIS-PDs with PN junctions maintained consistently low dark currents ($0.6\pm0.2$ pA) across detector lengths ranging from 7.5 to 100 \textmu m, approaching the detection limit of the measurement instrument ($\approx480$ fA). By contrast, dark currents of PIN-junction devices increased moderately with length, reaching $1.0\pm0.3$ pA for the longest device (100 \textmu m). Similarly, NIR-PDs with PN junctions exhibited dark currents $<$ 1 pA across all junction lengths (20 - 400 \textmu m), whereas PIN-junction devices showed increasing dark currents with length, reaching $5\pm0.5$ pA for 400-\textmu m-long PDs. 

\begin{figure} [t] 
    \centering
     \includegraphics[width=\textwidth]{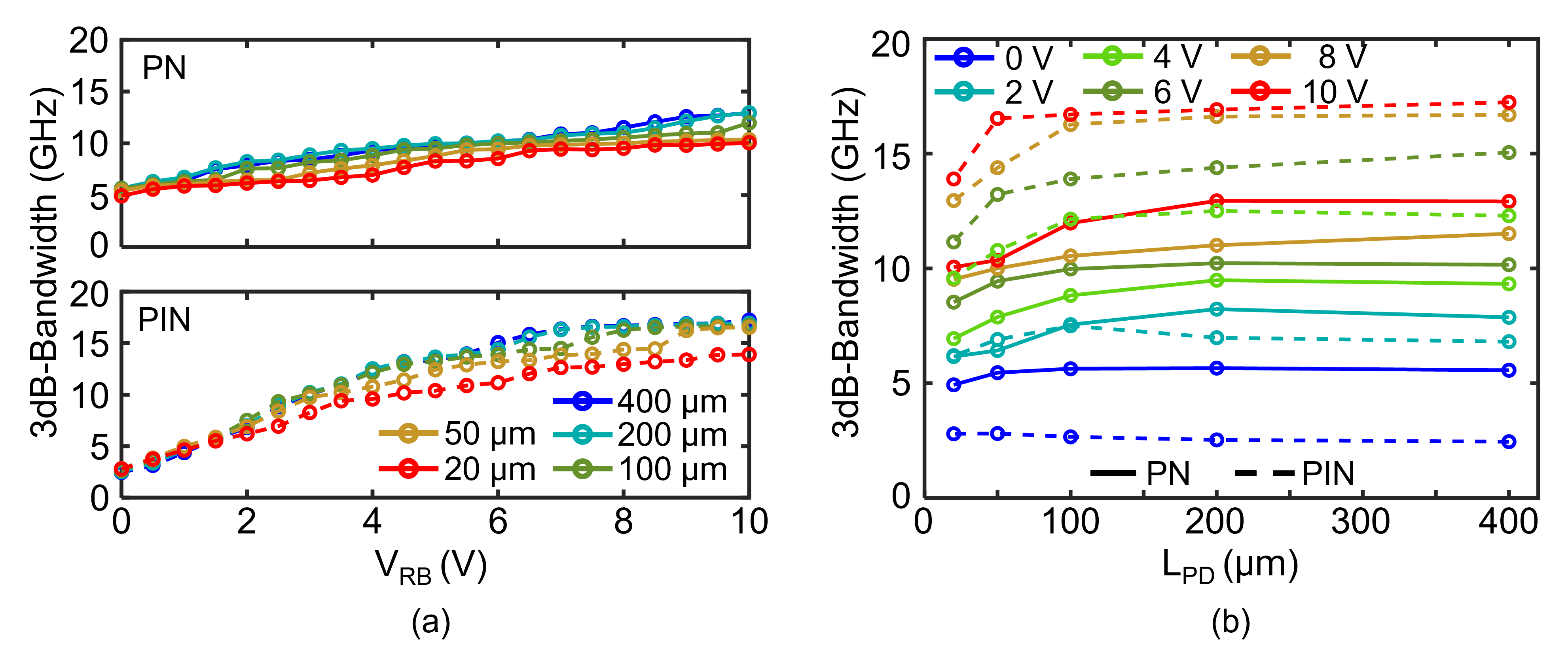}
    \caption{Frequency response characterization in linear mode. (a) Measured 3-dB optoelectronic bandwidth vs. reverse bias ($V_{RB}$) for NIR-PDs with varying $L_{PD}$ at an input wavelength of $\lambda$ = 785 nm ($P_{in,PD}$ = 226 \textmu W). (b) 3-dB bandwidth vs. $L_{PD}$ for different values of $V_{RB}$. Solid and dashed lines indicate PN- and PIN-junction devices, respectively. Raw frequency responses corresponding to this data are shown in Fig. \ref{fig:LinearMode_S21}. Each frequency response was smoothed using a Savitzky-Golay filter, and the 3-dB bandwidth was defined relative to the normalized response at 1 GHz.}
    \label{fig:Measurement_Bandwidth_LinearMode}
\end{figure}

\subsection{Optoelectronic bandwidth characterization}

Optoelectronic (OE) bandwidth measurements were performed for NIR-PDs (PN- and PIN-devices operating in linear mode; $V_{RB} \le10$ V) with varying $L_{PD}$, Fig. \ref{fig:Measurement_Bandwidth_LinearMode}. Corresponding raw and filtered radio frequency (RF) S21 responses are shown in Fig. \ref{fig:LinearMode_S21} (see Appendix). The 3-dB OE bandwidth generally increased with reverse bias, reaching up to 18 GHz (PIN, $V_{RB}=10$ V) and 13 GHz (PN, $V_{RB}$ = 10 V). The larger bandwidth of PIN-PDs is attributed to their wider depletion regions, set by the intrinsic region width, which result in lower junction capacitance compared to PN devices. In PIN-junction PDs, the bandwidth enhancement with increasing reverse bias is mainly driven by the increase in carrier drift velocity ($v_d$) with the electric field ($E$) in the depletion region ($v_d \propto E \propto V_{RB}$ \cite{quimby2006photonics}), continuing up to $\approx$8 V, beyond which the bandwidth saturates as $v_d$ approaches its saturation limit. This behavior indicates that PIN devices are predominantly transit-time limited, rather than limited by the resistor-capacitor (RC) time constant. For PN-PDs, 3-dB OE bandwidth also increases with reverse bias and follows an approximate square-root-dependence, $\propto\sqrt{V_{RB}+V_{bi}}$ (where $V_{bi}$ is the built-in voltage of the junction and measured to be 0.85 V). This response is attributed to the reduction in the junction capacitance with increasing reverse bias, consistent with the bias-voltage dependence of capacitance under the assumption of an abrupt doping profile \cite{wang2018temperature}. Thus, unlike PIN devices, PN-PDs are primarily RC-limited. Figure \ref{fig:Measurement_Bandwidth_LinearMode}(b) presents the 3-dB OE bandwidth as a function of $L_{PD}$ under varying reverse biases. For $L_{PD}>50$ \textmu m, the OE bandwidth remains largely independent of device length, confirming that the lateral junctions maintain the RC time constant and carrier transit time for varying longitudinal junction lengths. The observed reduction in bandwidth at shorter PD lengths ($L_{PD}=20$ \textmu m) is most likely due to lumped fringe capacitance ($\approx1-2$ fF \cite{piels201440}) becoming increasingly dominant as the junction length decreases \cite{piels201440,hosseini2007new}. Furthermore, the Via1 region is approximately 3\textmu m shorter than the Si junction along the longitudinal PD axis, possibly contributing significant capacitance and resistance as this length difference forms a larger fraction of $L_{PD}$.

\subsection{Avalanche-mode characterization}
\label{avalanche_gain_measurements}

In this subsection, we characterize PN-junction PDs operating in avalanche mode. The avalanche gain (multiplication factor), $M$, is a central figure-of-merit in avalanche-mode operation, defined relative to a unity-gain operating condition. In this work, the unity gain condition is established using PIN-junction PDs, which exhibit nearly constant photocurrent with increasing reverse bias ($V_{RB}\le20$ V). These devices are structurally similar to their PN-junction counterparts, differing only in the absence of moderately-doped regions used to define PN junctions. Specifically, the unity gain reverse bias ($V_{RB,M=1}$) is defined as the bias voltage at which the photocurrents of the PN- and PIN-PDs intersect. We calculate $M$ as 

\begin{equation}\label{eq:AvalancheGain}
    M(V_{RB}) = \frac{I_{ph}(V_{RB})-I_{dark}(V_{RB})}{I_{ph}(V_{RB,M=1})-I_{dark}(V_{RB,M=1})},
\end{equation}
where $I_{ph}(V_{RB,M=1})$ and $I_{dark}(V_{RB,M=1})$ are the photocurrent and dark current at unity gain, respectively. In the following measurements, dark current was averaged over five repeated measurements to suppress noise. In addition, due to the observed drift in I-V characteristics of the PDs when operating near avalanche breakdown (Fig. \ref{fig:IV_drifting}, see Appendix), the stabilization procedure described in Section \ref{section_fabrication} was applied prior to photocurrent and dark current measurements. 

\begin{figure} [t] 
    \centering
     \includegraphics[width=1\textwidth]{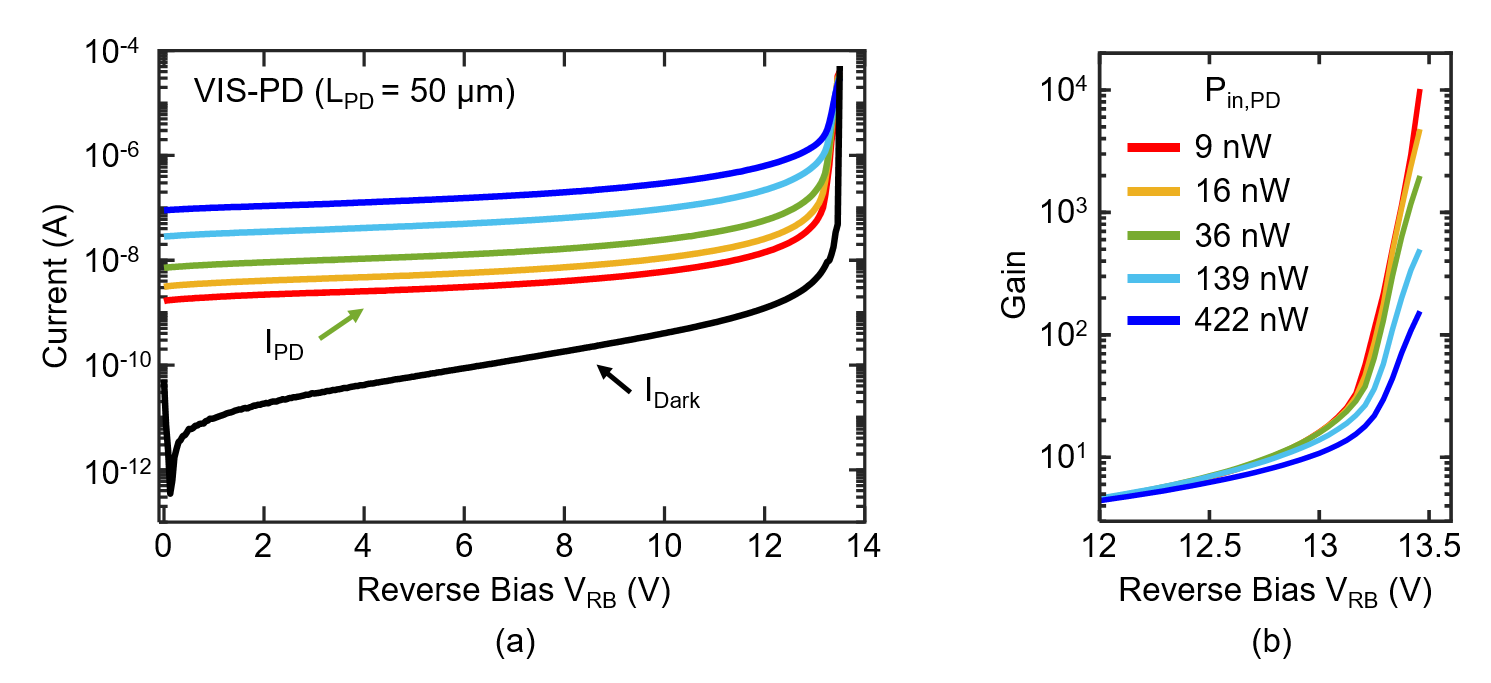}
    \caption{Avalanche gain measurements of a VIS-PD ($L_{PD} =$ 50 \textmu m). (a) Current vs. reverse bias voltage traces with different input optical powers ($P_{in,PD}$) ($\lambda=$ 532 nm) and under dark conditions. (b) Avalanche gain vs. reverse bias at different $P_{in,PD}$. Each calculated gain trace was averaged over five repeated measurements. Measurements were performed following the stabilization procedure described in Section \ref{section_fabrication}.}
    \label{fig:AvalancheGain_VIS}
\end{figure}

Figure \ref{fig:AvalancheGain_VIS}(a) shows the measured I-V traces of a 50-\textmu m-long VIS-PD.  The avalanche gains derived from these measurements are presented in Fig. \ref{fig:AvalancheGain_VIS}(b). The gain exhibited a strong dependence on the incident optical power, consistent with observations reported in Ref. \citenum{yanikgonul2021integrated}. Under low optical power conditions (9 nW), gains as high as $10^4$ were achieved, while higher optical input levels (422 nW) resulted in moderate gains on the order of 100. An additional PD was characterized across another nominally-identical photonic chip, with the gain varying by 14\% for optical input power of 422 nW at a 13.4-V reverse bias.

\begin{figure} [t] 
    \centering
     \includegraphics[width=1\textwidth]{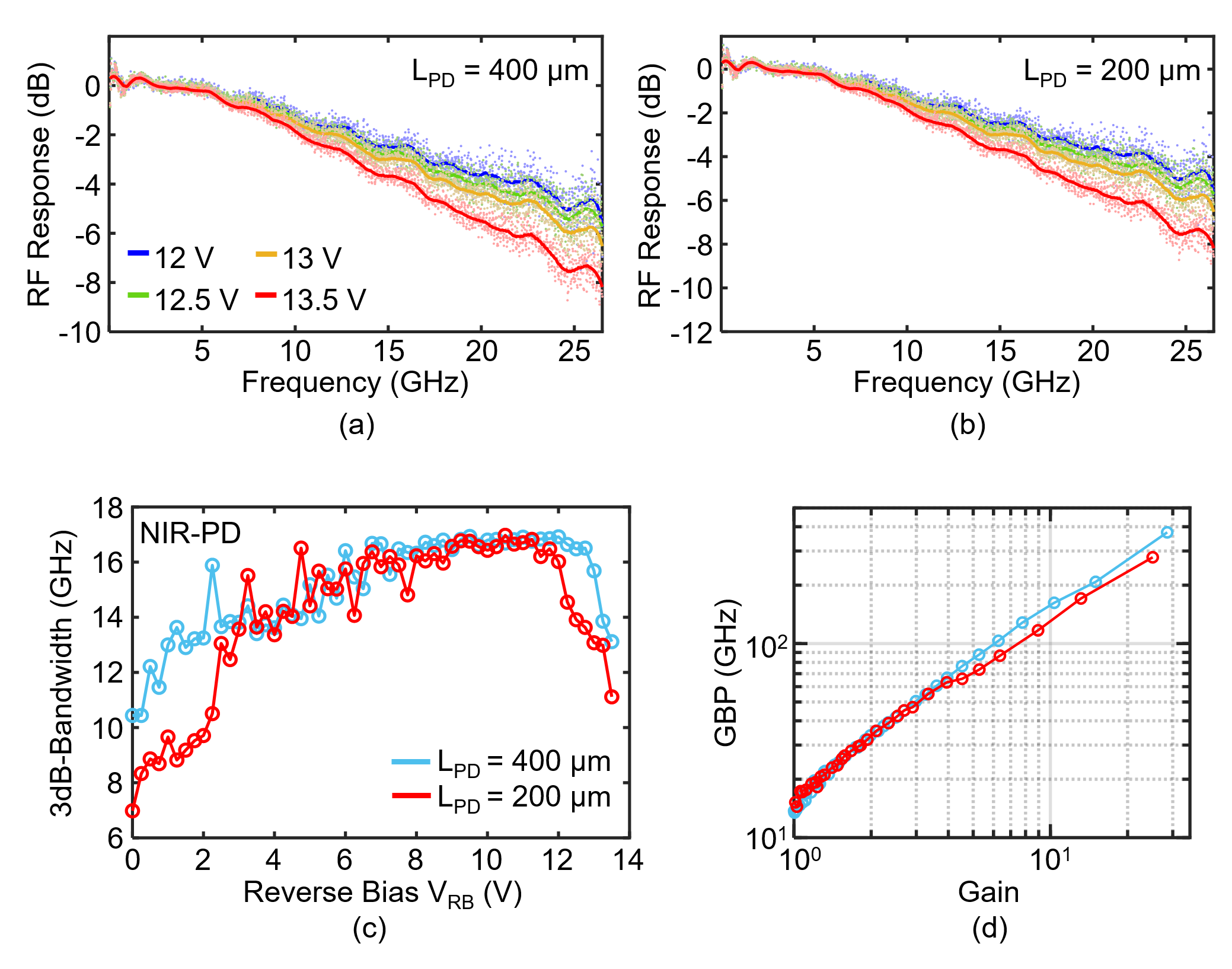}
    \caption{Avalanche-mode characterization of NIR-PDs. (a,b) Frequency response with PD lengths of (a) 400 \textmu m and (b) 200 \textmu m ($\lambda=$ 785 nm, $P_{in,PD}=$ 2.9 \textmu W); filtered data (solid lines), points (raw data). (c) Optoelectronic 3-dB bandwidth vs. reverse bias. (d) Gain-bandwidth product (GBP) vs. avalanche gain.}
    \label{fig:Bandwidth_GBP_avalanche}
\end{figure}

The RF responses of 200-\textmu m- and 400-\textmu m-long NIR-PDs operating in avalanche mode were also characterized, Fig. \ref{fig:Bandwidth_GBP_avalanche}. To ensure sufficient signal-to-noise ratios in the frequency response measurements, the optical input power (P$_{in,PD}$) was set to 2.9 \textmu W ($\lambda =$ 785 nm), corresponding to $M =$ 25 - 29 near the breakdown voltage. The measured RF responses at various bias voltages near breakdown are shown in Figs. \ref{fig:Bandwidth_GBP_avalanche}(a) and \ref{fig:Bandwidth_GBP_avalanche}(b). The resulting 3-dB OE bandwidths, Fig. \ref{fig:Bandwidth_GBP_avalanche}(c), initially increased with bias, similar to the behavior observed in linear-mode operation, but decreased near breakdown, an expected effect attributed to the avalanche build-up time \cite{emmons1967avalanche}. The corresponding maximum gain-bandwidth product (GBP), Fig. \ref{fig:Bandwidth_GBP_avalanche}(d), was calculated as 278 GHz ($M =$ 25) and 374 GHz ($M =$ 29) for $L_{PD} =$ 200 \textmu m and 400 \textmu m, respectively.

\section{Discussion}

\subsection{Length and polarization dependence}

External quantum efficiency (EQE) and dark current were measured for both VIS- and NIR-PDs with different device lengths, Fig. \ref{fig:Measurement_EQEs}. The observed length dependencies suggest guidelines for design of this device dimension. For VIS-PDs, the EQEs saturate beyond a PD length of 50 \textmu m, indicating that additional length provides negligible enhancement in responsivity. By contrast, NIR-PDs show a gradual EQE increase with length; however, this enhancement is modest, e.g., going from 200 \textmu m to 400 \textmu m yields $<20\%$ increase in EQE outside the hybridization wavelengths ($\lambda=$ 650-700 nm). Dark current measurements, in contrast, reveal a substantial length dependence in both variants, particularly in PIN-junction devices, Figs. \ref{fig:Measurement_EQEs}(d) and \ref{fig:Measurement_EQEs}(e). In VIS-PDs (PIN), increasing the junction length from 50 \textmu m to 100 \textmu m nearly doubles the dark current. Similarly, in NIR-PDs (PIN), dark current increases from $\approx200$ pA to $>$ 500 pA as the length increases from 200 \textmu m to 400 \textmu m. In comparison, PN-junction devices maintain considerably lower dark currents  (consistently $<1$ pA across all tested lengths), highlighting a distinct advantage for low-noise operation. The higher dark current in PIN-junction devices is attributed primarily to the wider intrinsic region (2 \textmu m), which corresponds to a larger volume for thermal carrier generation and an increased surface area for Shockley–Read–Hall (SRH) generation from traps at the Si/SiO$_2$ boundaries \cite{chen2016dark}. Overall, these results demonstrate a trade-off: while increasing the device length can improve responsivity up to the point set by the coupling to and absorption in the Si junction [Figs. \ref{fig:Schematic_PD}(c) and \ref{fig:Schematic_PD}(d)], it also increases dark current, particularly in PIN-junction devices, which in turn degrades the signal-to-noise ratio. Lengths of 50 \textmu m for VIS-PDs and 200 \textmu m for NIR-PDs represent a compromise, achieving high EQE without excessively increasing dark current. Devices with these lengths also exhibited highly linear optical responses over a broad input-power range, achieving dynamic ranges $>50$ dB with excellent linearity.

While the measurements in this work focus on TE-polarization device operation, the PDs are also capable of efficiently detecting TM-polarized light, and they exhibited broad wavelength spans with low polarization dependence. Additional measurements in Fig. \ref{fig:Measurement_EQEs_TM} (see Appendix) show EQE spectra for TM-polarized light at a 2-V reverse bias, matching the TE polarization test conditions. For the VIS-PDs, data from six nominally identical photonic chips show that 50-\textmu m-long devices achieved an average EQE $>52\%$ over $\lambda=400-550$ nm. Similar to the TE polarization case, EQE rolled off toward longer wavelengths due to the reduced Si absorption, while near-complete absorption was observed at shorter (blue) wavelengths. Compared to TE polarization operation, EQE increased only modestly with length for TM input, by $\lesssim20\%$ from 20 \textmu m to 100 \textmu m. Also, the hybridization-related localized spectral peaks were much less pronounced for TM-polarized light, Fig. \ref{fig:Measurement_EQEs_TM}(b). For $L_{PD} = 50$ \textmu m, the ratio of TE- and TM-polarization average EQE spectra showed a low polarization dependence of $<0.48$ dB at blue/green wavelengths ($\lambda = 400-532$ nm), increasing to $<2$ dB over $\lambda = 400-560$ nm. For NIR-PDs with TM input, the measured average EQE exceeded 70\% across $\lambda$ = 600-700 nm ($L_{PD}\ge200$ \textmu m) over seven chips, indicating strong absorption at red wavelengths, Fig. \ref{fig:Measurement_EQEs_TM}(c). EQE remained $> 60\%$ up to $\lambda=$ 800 nm and then decreased steadily, reaching an average EQE of $\approx16\%$ at $\lambda$ = 955 nm. Length had a stronger impact here than in VIS-PDs: EQE increased from $\approx20\%$ at $L_{PD}$ = 20 \textmu m to $>$80\% at $L_{PD}=200$ \textmu m over $\lambda=700-800$ nm. Consistent with TE results, 200-\textmu m-long and 400-\textmu m-long devices exhibited closely matched EQEs (within 10\%). The EQE polarization dependence of the NIR-PD ($L_{PD}=200$ \textmu m) was $<1.4$ dB over $\lambda=600-748$ nm and remained $<3$ dB across the full measurement range of $\lambda=560-955$ nm.

\subsection{Optoelectronic bandwidth limits}

The optoelectronic bandwidth characterization of the PIN- and PN-junction devices provides insights into the underlying mechanisms. In PIN devices, the depletion region is defined by the intrinsic region (width of 2 \textmu m) and is nearly bias-independent \cite{quimby2006photonics}. Using a parallel-plate capacitor model, the junction capacitance ($C_j$) per length was estimated to be 0.0114 fF/\textmu m. Combining this with the measured resistance per unit length, 5.6 k$\Omega\cdot$\textmu m (from the slope of the forward-bias I-V trace), gives a length-independent RC that corresponds to an RC-limited bandwidth in the THz range. This rules out RC limitation and explains the lack of length-dependence in the measured bandwidth. However, the 3-dB bandwidth ($f_{3dB}$) consistently increases with reverse bias ($V_{RB}$) up to $\approx8$ V, Fig. \ref{fig:Measurement_Bandwidth_LinearMode}(a), which we attribute to the increase in carrier drift velocity with electric field and the consequent reduction of transit time. For {$V_{RB}\gtrsim$ 8 V, the bandwidth shows little further increase as the carrier drift velocity saturates, due to bias-independent charge mobility under high electric fields \cite{quimby2006photonics}. These observations indicate that PIN devices are predominantly transit-time limited over the measured bias range. 

In PN-PDs, the depletion width ($W_{dep}$) expands with reverse bias while the carrier drift velocity also increases due to the stronger electric field. Since both the transit path and drift velocity scale in opposing ways, the net transit time remains approximately bias-independent \cite{quimby2006photonics}. Thus, the observed bandwidth increase is mainly attributed to the bias-dependent junction capacitance. Specifically, for $V_{RB}\gtrsim5$ V the measured 3-dB bandwidth follows an approximate square-root dependence on bias, consistent with a PN junction with an abrupt doping profile where $f_{3dB}\propto C_{j}^{-1}\propto \sqrt{V_{RB}+V_{bi}}$ \cite{wang2018temperature}. At lower biases, deviations from this square-root dependence were observed, which we attribute to the graded impurity transition across the doped junction. When the narrower depletion region resides largely within this transition region, the depletion-width dependence becomes weaker, approaching $W_{dep}\propto(V_{RB}+V_{bi})^{1/3}$, yielding $C_j\propto(V_{RB}+V_{bi})^{-1/3}$ and corresponding bias dependence of $f_{3dB}$ \cite{sze2021physics}. Technology Computer-Aided Design (TCAD) cross-section simulations of the doping and space-charge distributions verify the graded transition of net doping density nearby the junction center and change in the charge distributions for different biases, Fig. \ref{fig:TCADresults} (see Appendix). These results support that PN devices are predominantly RC-limited, with their bandwidth determined by the bias-dependent junction capacitance rather than transit time.

\subsection{Avalanche-mode bandwidth and drift}

As the reverse bias approaches breakdown and avalanche multiplication becomes prominent, a reduction in optoelectronic bandwidth was observed Fig. \ref{fig:Bandwidth_GBP_avalanche}(c). For NIR-PDs with $L_{PD}$ = 200 \textmu m and 400 \textmu m, the bandwidth dropped from $\approx16$ GHz to 11 GHz and 13 GHz, respectively, when $V_{RB}\ge12$ V. While the initial increase in bandwidth at moderate biases is primarily attributed to reduced junction capacitance, the subsequent roll-off at higher biases is consistent with avalanche-related effects, particularly the impact-ionization build-up time (carrier multiplication delay) \cite{emmons1967avalanche}. These results indicate a bias-dependent shift in the dominant limitation for optoelectronic bandwidth in PN-type devices: capacitance-limited regime at low-moderate reverse bias, transitioning to avalanche-induced bandwidth degradation at high reverse bias. Despite the observed bandwidth reduction in avalanche regime, the substantial increase in gain yields a continued enhancement in gain-bandwidth product (GBP); at a 13.5-V reverse bias, GBP reached 278 GHz and 374 GHz for the 200-\textmu m and 400-\textmu m NIR-PDs, respectively, Fig. \ref{fig:Bandwidth_GBP_avalanche}(d). 

While the avalanche-mode results highlight high gains and GBP, operating in this high-field regime also introduces non-idealities that impact measurement repeatability. In particular, near breakdown we observed a drift in the current-voltage (I-V) characteristics, including the breakdown voltage and measured photo/dark currents, between successive bias voltage sweeps, [Figs. \ref{fig:IV_drifting}(a) and \ref{fig:IV_drifting}(b); see Appendix]. This drift may arise from several possible mechanisms. Prior avalanche photodetector studies have attributed I-V drifts to trap charging in passivation/oxide and at Si/SiO$_2$ interfaces \cite{gurtler2005avalanche,saraswat1978breakdown,nicollian1971electrochemical}, redistribution of mobile surface charges \cite{atalla1959stability,shockley1964mobile}, and humidity-dependent surface conductivity \cite{beck2001breakdown}. Thus, we attempted to suppress the drift by storing the test chips in dry nitrogen and performing the I-V characterization at low humidity. Despite these precautions, the drift persisted, indicating that ambient air is not the sole factor. We hypothesize that intrinsic device-level effects introduced during fabrication, such as fixed surface charges and interface trap states in the passivation oxide, dominate: under high field near breakdown these states can (dis)charge, gradually reshaping the junction electrostatics and shifting the effective breakdown. To reduce the variability in measurements, we adopted a pre-biasing protocol (holding high bias followed by voltage sweeps up to a lower final bias), which partially stabilized the I-V response but did not completely eliminate the drift, Fig. \ref{fig:IV_drifting}(c). Further investigations are needed to isolate the dominant mechanism, and fabrication-level approaches (e.g., improved interface quality, surface-field control \cite{beck2001breakdown}) may help suppress or compensate for long-term charge migration in subsequent device generations.

\subsection{Performance comparison and outlook}

Compared to previously reported waveguide-coupled photodetectors operating at submicrometer wavelengths (VIS–NIR) (Table \ref{tab:comparison_linear}; see Appendix), our SiN-on-SOI PD variants offer among the lowest reported dark currents in linear mode, remaining below 1 pA (VIS-PD) and 2 pA (NIR-PD) at a 2-V reverse bias, without compromising responsivity. Only the SiN end-coupled Si PD in Ref. \citenum{yanikgonul2021integrated} achieved comparable pA-scale dark currents. We report EQEs $>$60\% across broad continuous spectral bands (400-560 nm for the VIS-PD and 600-748 nm for the NIR-PD) with peak values exceeding 84\%. These EQEs are comparable to the highest reported for short-wavelength monolithically integrated PDs, while the combined optical bandwidth of both variants extends beyond prior demonstrations, which were limited to narrower ranges and characterization at discrete wavelengths \cite{lin2022monolithically,morgan2021waveguide, de2022amorphous,yanikgonul2021integrated, cuyvers2022heterogeneous}. Optoelectronic bandwidths up to 18 GHz (measured at $\lambda = 785$ nm) place our devices among the fastest linear-mode short-wavelength PDs reported in this range, with only three faster demonstrations: an end-coupled Si PD \cite{yanikgonul2021integrated} (30 GHz), a photon-trapping Si PD \cite{gao2017photon} (19.5 GHz), and a GaAs-based metal–semiconductor–metal PD \cite{chen2018integration} (20 GHz). Additionally, similar bandwidths are sustained across multiple device lengths ($L_{PD}>$ 50 \textmu m), Fig. \ref{fig:Measurement_Bandwidth_LinearMode}(b). However, at longer wavelengths (800-955 nm), our reported EQEs are more modest, $35\%$ at 800 nm and $23\%$ at 850 nm, relative to structures that employ photon-trapping \cite{gao2017photon}, grating-assisted coupling into Si \cite{pour2017high} or a tapered Si rib structure beneath the SiN waveguide \cite{chatterjee2019high}. Substantial EQEs remain feasible via geometry refinements of the SiN waveguide and Si junction to increase absorption. Future investigations aimed at improving mode matching between the SiN guided mode and higher-order Si slab modes --- as already observed across 400–748 nm in our devices [Fig. \ref{fig:ModeSimulations_VISPD_NIRPD}(d)] --- may enable hybridization at longer wavelengths and extend high-EQE performance deeper into the NIR.

In comparison with previous demonstrations of submicrometer-wavelength waveguide-coupled avalanche Si photodetectors (APDs) on integrated platforms (Table \ref{tab:comparison_APD}), our devices achieve among the highest reported gain–bandwidth products (GBPs): 278 GHz for $L_{PD}=200$ \textmu m and 374 GHz for $L_{PD}=400$ \textmu m, while maintaining low dark currents ($<20$ nA). The NIR-PDs exhibited optoelectronic bandwidths of 10 and 13 GHz, values comparable to prior demonstrations, though exceeded by the end-coupled Si PD in Ref. \citenum{yanikgonul2021integrated} (19 GHz) and the grating-assisted PD in Ref. \citenum{pour2017high} (16.4 GHz). At $\lambda = 785$ nm and with an input optical power of 2.9 \textmu W, we measured avalanche gains of 25 (200 \textmu m) and 29 (400 \textmu m). Since avalanche multiplication depends strongly on input power, direct comparisons across studies are challenging. For example, comparable gains ($M \approx $ 6–11) have been reported, but at substantially higher input levels, e.g., 0.57 mW \cite{yanikgonul2022high} and $\approx 1$ mW \cite{pour2017high}, with the expectation that larger gains could be achieved with lower input optical powers. In our devices, higher optical powers were not explored due to increasing drift observed in the avalanche-mode I–V characteristics at elevated input levels. Consistent with this power-dependent behavior, our VIS-PDs ($L_{PD}=50$ \textmu m) at $\lambda = 532$ nm demonstrated gains up to $10^4$ at an input of $\approx 10$ nW, highlighting the potential for strong multiplication in the low-light regime at NIR wavelengths as well. Although bandwidth measurements were performed at $\lambda = 785$ nm, the combination of high GBPs and avalanche-mode bandwidths up to 13 GHz at this wavelength, together with broadband absorption and EQE characteristics, suggests that high-speed, high-gain operation is feasible across the full wavelength range of these devices. To further enhance the avalanche-mode bandwidth and overall GBP, several design improvements can be pursued, including engineering the doping profile \cite{yanikgonul2021integrated}, reducing the width of the multiplication region to minimize transit time and impact-ionization delay \cite{campbell2015recent}, and incorporating photonic enhancement structures such as photon-trapping layers or resonant cavities \cite{zang2017silicon}. These modifications may enable even higher-speed, low-noise operation while maintaining high avalanche gains and compatibility with foundry-fabricated SiN-on-SOI platforms.

These attributes are advantageous in short-wavelength systems requiring fast, low-noise detection. Examples include biophotonics applications such as optical coherence tomography (OCT) around 800 nm \cite{rank2021toward}, label-free biomolecule sensing at 850 nm \cite{melnik2016local}, and enzyme-absorption assays near 650 nm \cite{nitkowski2011chip}. Applications also arise in programmable and quantum photonic circuits, where high–signal-to-noise detection across a broad wavelength range can enable feedback control of large networks of tunable interferometric devices with flexibility in operating wavelength (e.g., for multi-wavelength state multiplexing) \cite{dong2022high, palm2023modular, puckett2021422, bogaerts2020programmable, arrazola2021quantum}. Further applications include low-light fluorescence detection and on-chip power monitoring in photonic integrated circuits for optogenetics and neurotechnology \cite{archetti2019waveguide, diekmann2017chip, roszko2025foundry, Mu2025Nanophotonic_CLEO}, as well as blue/green underwater optical communication \cite{wu2017blue, notaros2023liquid}.

\section{Conclusion}

In conclusion, we have demonstrated broadband, monolithically integrated, waveguide-coupled photodetectors within a submicrometer-wavelength, foundry-fabricated Si photonics platform. Each device features a SiN waveguide passing over top a doped Si detector, exciting hybrid modes shared between the layers. Engineering of the SiN waveguide dimensions determines coupling into these hybrid modes and their respective absorption characteristics. Building on this principle, we designed and experimentally demonstrated two detector variants that collectively span wavelength ranges of 400 - 748 nm and 749 - 955 nm with external quantum efficiencies exceeding 60\% and 12\%, respectively. The devices further exhibited low dark currents $<$ 2 pA (with 2 V reverse bias), linear dynamic ranges $>$ 50 dB, optoelectronic bandwidths up to 18 GHz, and avalanche operation with a gain-bandwidth product of 374 GHz. Overall, this work introduces a new building block to the growing library of integrated photonic components for submicrometer wavelengths. Spanning both visible and near-infrared wavelengths, and offering operational flexibility through junction design and reverse bias control --- including high-dynamic-range linear, high-speed, and avalanche detection modes --- the photodetectors demonstrated here open new avenues for realizing scalable PICs operating at short wavelengths. Ongoing work aims to enhance detection efficiency for wavelengths between 800 and 1000 nm through further optical engineering, and to characterize device performance in Geiger mode. %
\section*{Appendix}

In this appendix, we provide additional simulation and characterization results for our waveguide-coupled PDs and, also, compare the device performance to previously reported results. The simulated doping profile and space-charge distributions for PN-junction PDs are shown in Fig. \ref{fig:TCADresults}, obtained from Technology Computer-Aided Design (TCAD) tools (Synopsys Sentaurus). Figure \ref{fig:FDTD_comparison_WaveguideGeometry} presents simulations of PD absorption with alternative SiN thicknesses, showing visible-spectrum performance for the full-thickness SiN1 layer and red/near-infrared performance for the partially-etched SiN1p layer. The following additional characterization results are presented: EQE measurements for TM-polarized light (Fig. \ref{fig:Measurement_EQEs_TM}), RF responses of NIR-PDs operating in linear mode (Fig. \ref{fig:LinearMode_S21}), and measurements of drift in PD current-voltage characteristics during avalanche-mode operation (Fig. \ref{fig:IV_drifting}). Additionally, in Table \ref{tab:comparison_linear}, we compare the performance of on-chip photodetectors for submicrometer-wavelength light operating in linear mode (i.e., unity- or low-gain operation). Table \ref{tab:comparison_APD} compares the performance of integrated photodetectors with avalanche or photoconductive gain.

\begin{figure} [b] 
    \centering
     \includegraphics[width=\textwidth]{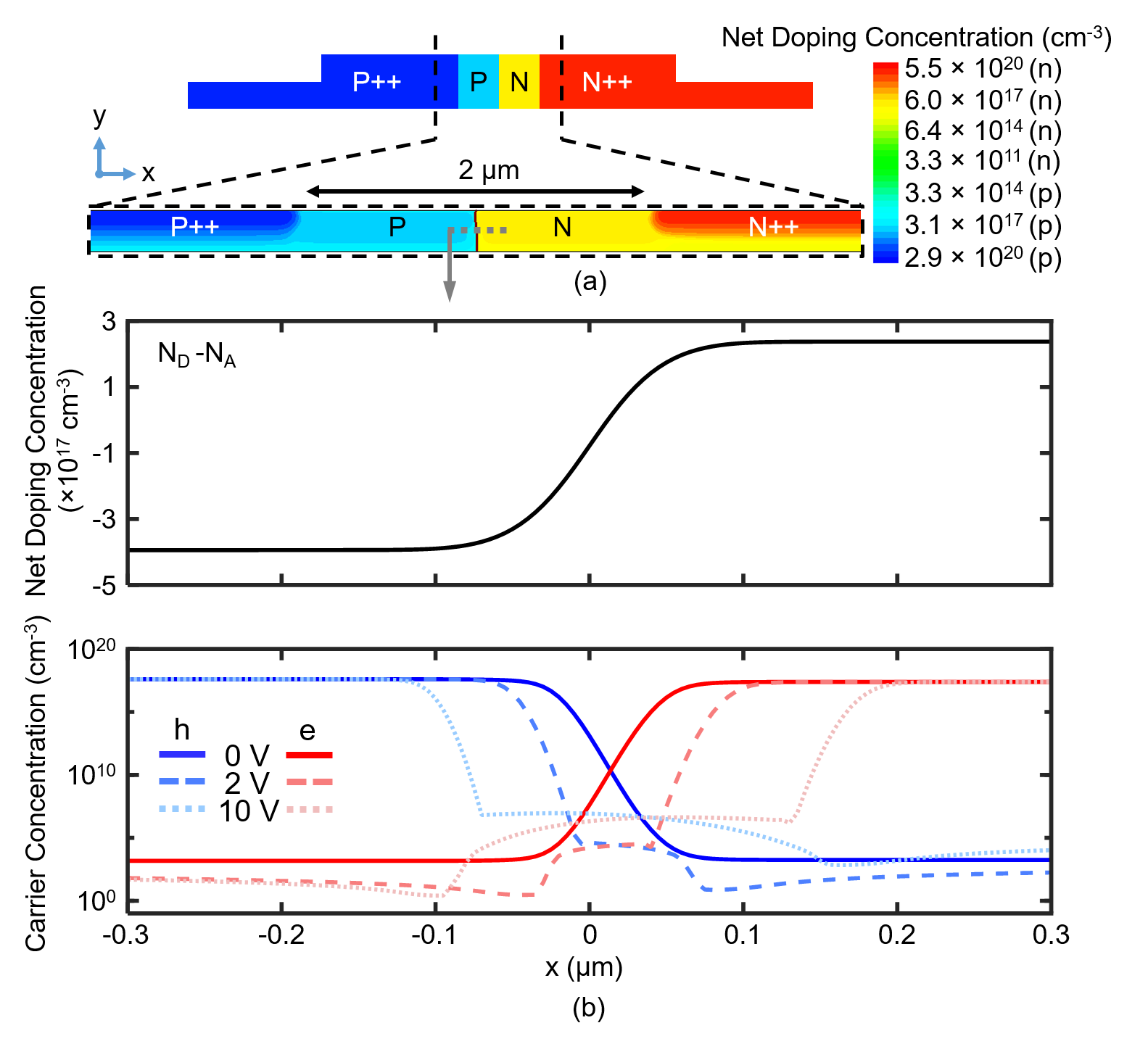}
    \caption{Simulated cross-sectional doping and carrier profiles. (a) Net doping concentration of the PN junction cross-section, defined as the donor concentration ($N_D$) minus the acceptor concentration ($N_A$), i.e., $N_D-N_A$. (b) Net doping concentration (top) and free-carrier concentrations (bottom) along a cutline indicated by the grey dashed line, centered vertically in the Si and extending laterally $\pm$30 nm around the junction center. Electron (e) and hole (h) distributions are shown for reverse biases of 0, 2, and 10 V.}
    \label{fig:TCADresults}
\end{figure}
\clearpage

\begin{figure} [t] 
    \centering
     \includegraphics[width=0.9\textwidth]{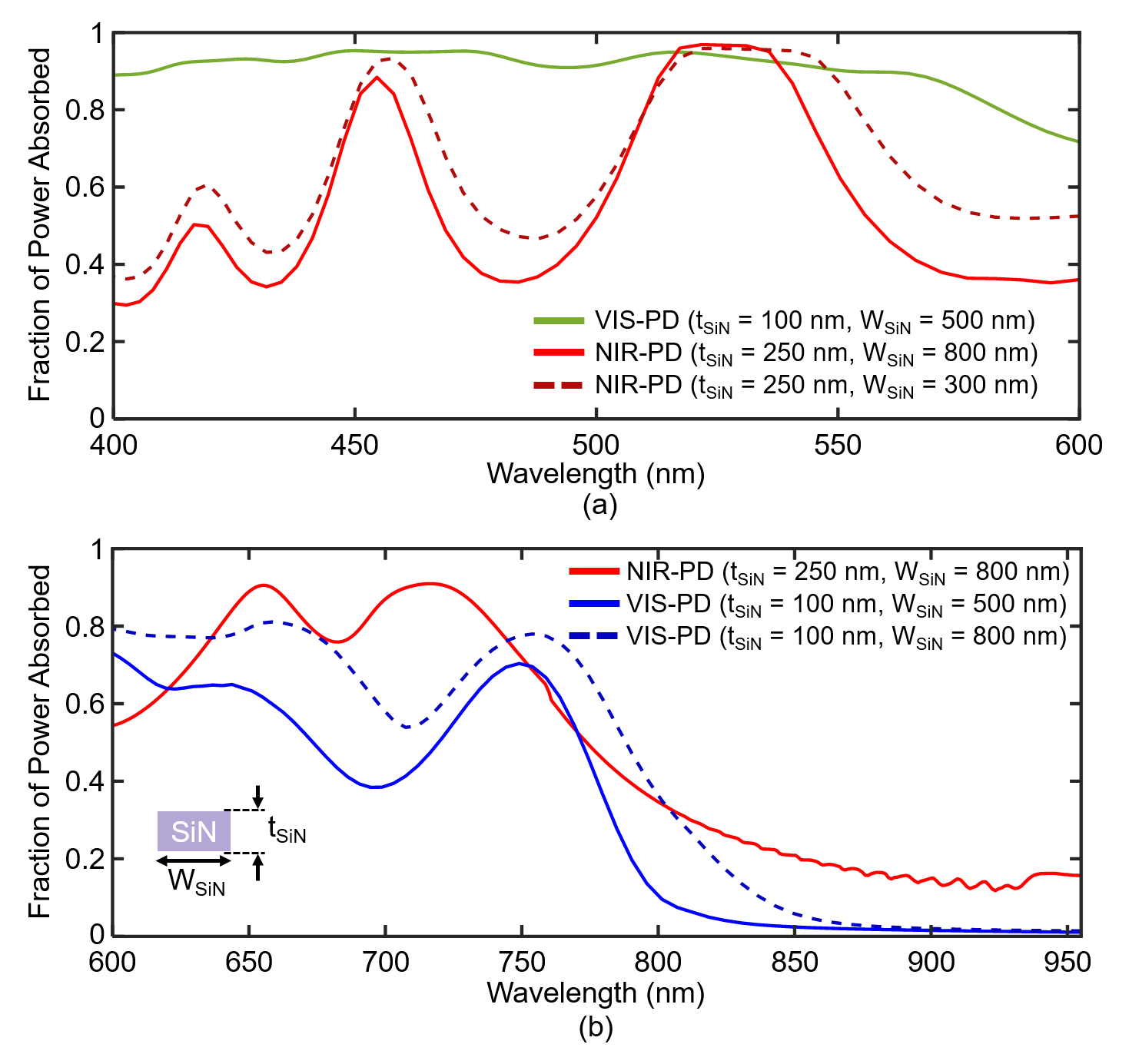}
    \caption{Simulated PD absorption with alternative SiN thicknesses. (a) Visible-spectrum absorption for PDs with full-thickness SiN1 (250 nm); $L_{PD} = $ 100 \textmu m. The absorbed optical power fraction versus wavelength is shown for the VIS-PD design (reference, green solid trace), the NIR-PD design (red solid trace), and a modified NIR-PD design with a narrower SiN1 waveguide (red dashed trace). (b) Red/near-infrared absorption for PDs with the partially-etched SiN1p layer (100 nm); $L_{PD} = $ 200 \textmu m. The absorbed power fraction is shown for the NIR-PD design (reference, red solid trace), the VIS-PD design (blue solid trace), and a modified VIS-PD design with a wider SiN1p waveguide (blue dashed trace). $t_{SiN}$: SiN thickness, $W_{SiN}$: SiN waveguide width.}
    \label{fig:FDTD_comparison_WaveguideGeometry}
\end{figure}

\begin{figure} [b] 
    \centering
     \includegraphics[width=0.94\textwidth]{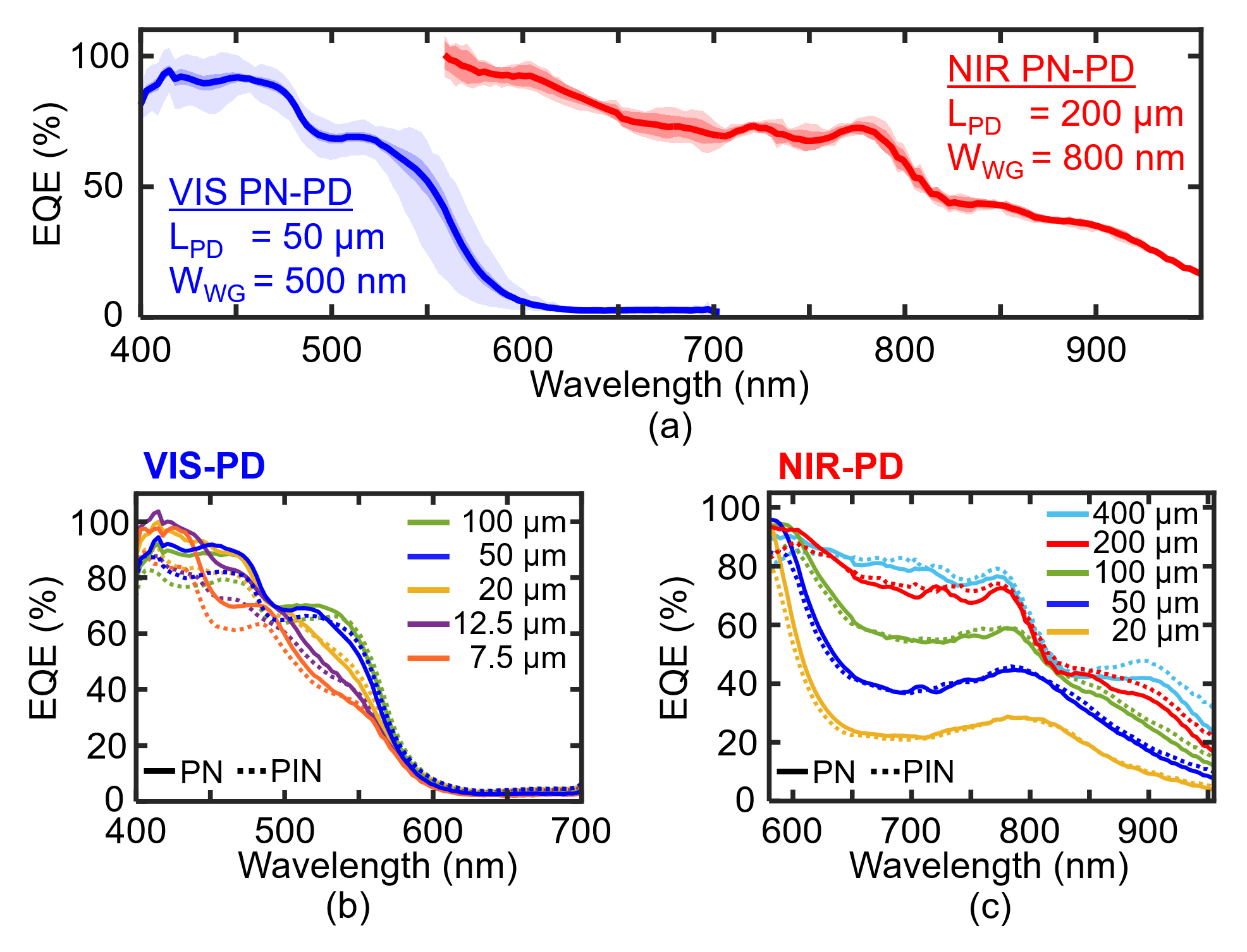}
    \caption{Linear-mode EQE characterization of PDs with TM-polarized input light. (a) Measured EQE spectra for VIS-PDs (6 chips) and NIR-PDs (7 chips). Solid lines indicate the average EQE across chips, inner shaded regions represent the standard error, and outer shaded regions denote the range (maximum/minimum) of the measured values. (b,c) Average EQE spectra of (b) VIS-PDs and (c) NIR-PDs vs. wavelength for different PD lengths ($L_{PD}$); PN-junction and PIN-junction devices are represented with solid and dashed lines, respectively.}
    \label{fig:Measurement_EQEs_TM}
\end{figure}

\begin{figure} [!t] 
    \centering
     \includegraphics[width=\textwidth]{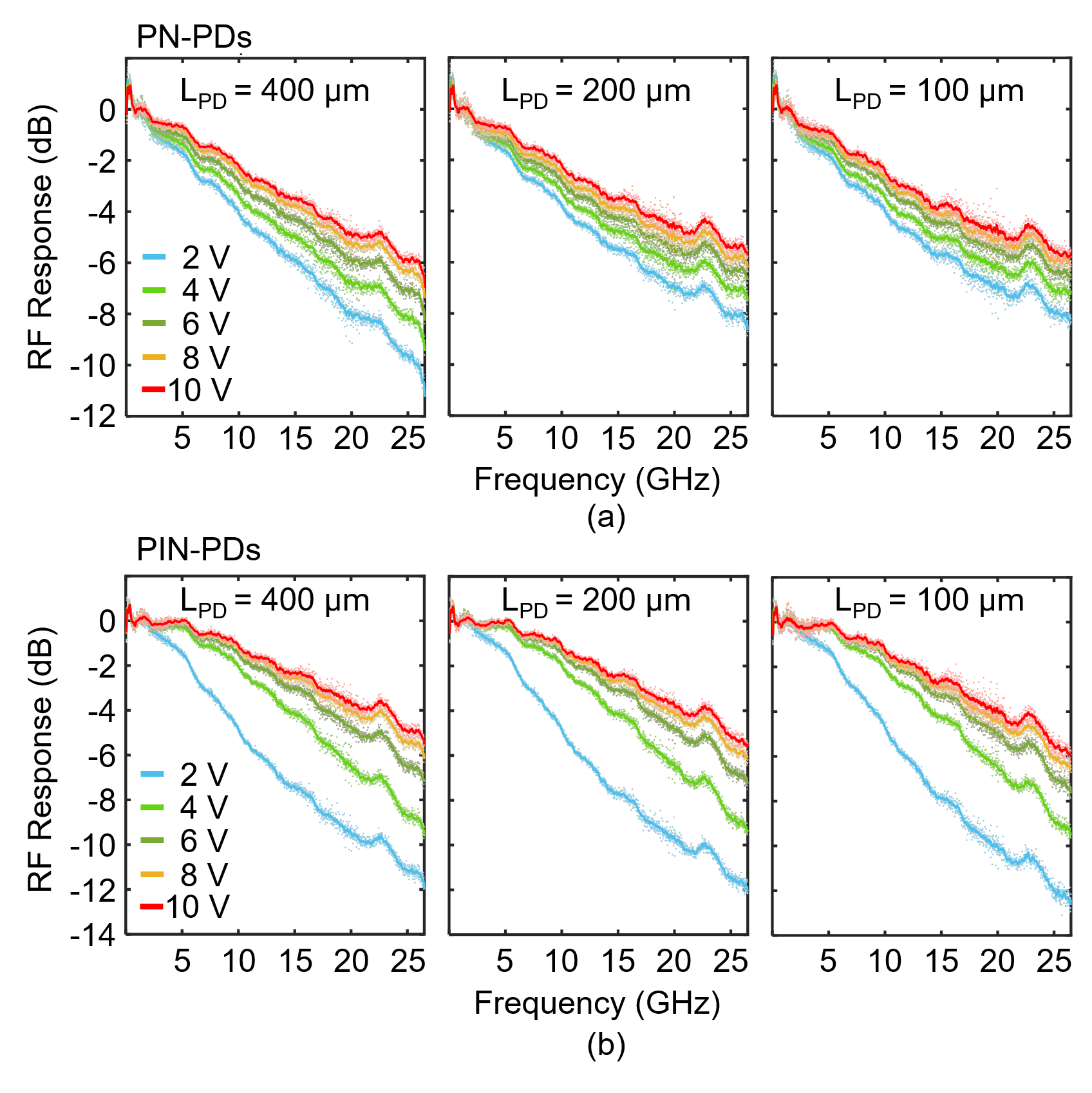}
    \caption{Optoelectronic bandwidth measurements of (a) PN- and (b) PIN-junction NIR-PDs of various lengths operating in linear mode; $\lambda$ = 785 nm, $P_{in,PD}=$ 226 \textmu W,  $V_{R}\le$ 10 V. Filtered data: solid lines, raw data: points.}
    \label{fig:LinearMode_S21}
\end{figure}
\clearpage

\begin{figure} [!t] 
    \centering
     \includegraphics[width=1\textwidth]{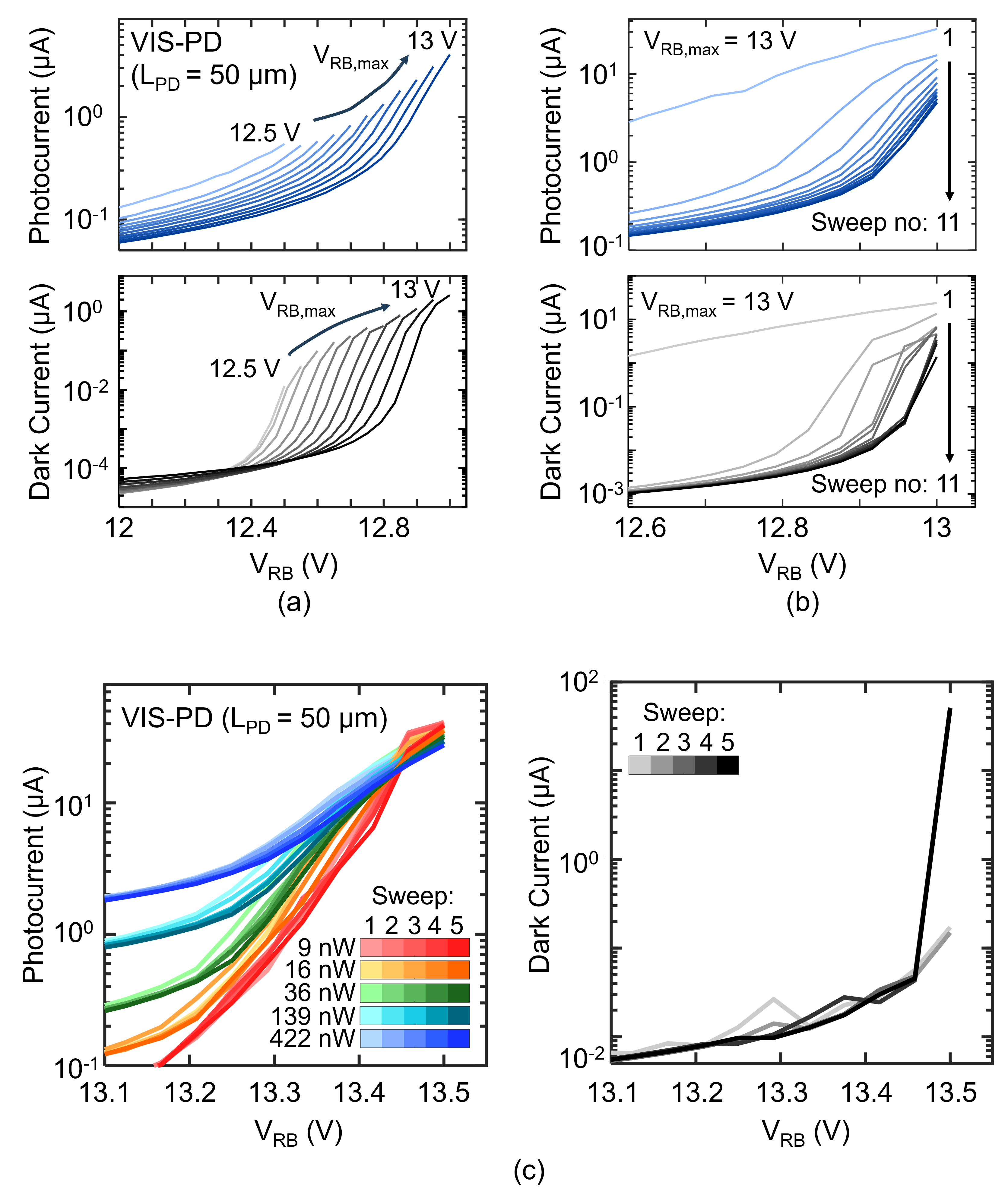}
    \caption{Variation of PD current-voltage characteristics during avalanche gain measurements; VIS-PD, $L_{PD} =$ 50 \textmu m. (a) Sequence of 11 I-V traces with optical input (top, $\lambda=$ 532 nm, $P_{in,PD}=35$ nW) and in dark conditions (bottom). The maximum reverse bias was progressively increased across the measured traces. (b) Sequence of 11 I-V traces of photocurrent and dark current for a nominally identical device to that of (a). The maximum reverse bias was fixed at 13 V for all traces. (c) I-V characterization of the VIS-PD in avalanche mode with varying optical input powers (left) and in dark conditions (right). Five sequential I-V traces were acquired in each condition. Calculated avalanche gains from (c) are shown in Fig. \ref{fig:AvalancheGain_VIS}.}
    \label{fig:IV_drifting}
\end{figure}
\clearpage

\begin{table}[!t]
\centering
\caption{Comparison of submicrometer-wavelength (VIS and NIR) integrated photodetectors operating in linear mode}
\label{tab:comparison_linear}
\resizebox{\textwidth}{!}{%
\begin{tabular}{ccccccc}
\hline
\textbf{Ref.} &
  \textbf{Type} &
  \textbf{Wavelength} &
  \textbf{$V_{RB}$} &
  \textbf{$I_{dark}$} &
  \textbf{EQE} &
  \textbf{BW}  \\ \hline
This work &
 \begin{tabular}[c]{@{}c@{}}SiN-on-SOI (VIS-PD) \end{tabular} & 
 400-600 nm &
 \begin{tabular}[c]{@{}c@{}}2 V\end{tabular} &
 \begin{tabular}[c]{@{}c@{}}$<$1 pA\end{tabular} &
 \begin{tabular}[c]{@{}c@{}}$>61\%$$^f$ \\ $61\%-24\%$$^g$ \end{tabular}  &
 \begin{tabular}[c]{@{}c@{}}N.A\end{tabular} \\
This work &
 \begin{tabular}[c]{@{}c@{}}SiN-on-SOI (NIR-PD)
 \end{tabular} & 
 600-955 nm &
 \begin{tabular}[c]{@{}c@{}}2 V\end{tabular} &
 \begin{tabular}[c]{@{}c@{}}$<$2 pA\end{tabular} &
 \begin{tabular}[c]{@{}c@{}}$>$60$\%$$^h$ \\ $60\%-12\%$$^i$\end{tabular} &
 \begin{tabular}[c]{@{}c@{}} up to 18 GHz\end{tabular} \\
\cite{lin2022monolithically} &
 \begin{tabular}[c]{@{}c@{}}SiN-on-Si bulk \\ evanescently coupled \end{tabular} &
  405-640 nm$^a$ &
  2 V &
  144$\pm$42 pA$^e$ &
  $>$ 60$\%$$^j$ &
  4.4$\pm$1.1 GHz \\
\cite{morgan2021waveguide} &
  \begin{tabular}[c]{@{}c@{}}Al$_2$O$_3$-on-Si \\ evanescently coupled
  \end{tabular} &
  405 nm &
  2 V$^b$ &
  $<$ 1 nA &
  73$\%$$^b$ &
  N.A. \\
\cite{de2022amorphous} &
  \begin{tabular}[c]{@{}c@{}}Integrated amorphous \\Si PD \end{tabular} &
  \begin{tabular}[c]{@{}c@{}}  660 nm \end{tabular} &  
  8 V &
  50 pA &
  $5.6\%$$^b$ &
  N.A. \\   
\cite{yanikgonul2021integrated} &
  \begin{tabular}[c]{@{}c@{}} SiN end-coupled (lateral doping) \\ SiN end-coupled (interdigitated doping)
  \end{tabular} &
  685 nm &
  2 V &
  \begin{tabular}[c]{@{}c@{}}  $\approx$1 pA \\ $<70$ pA
  \end{tabular} &
  \begin{tabular}[c]{@{}c@{}}  $>85\%$$^{b,k}$
  \end{tabular} &
  \begin{tabular}[c]{@{}c@{}}  up to 30 GHz \\ up to 18 GHz$^b$
  \end{tabular}\\
\cite{cuyvers2022heterogeneous} &
  \begin{tabular}[c]{@{}c@{}}Heterogeneously integrated \\Si PD \end{tabular} &
  \begin{tabular}[c]{@{}c@{}}  775 nm \\ 800 nm \end{tabular} &  
  3 V &
  107 pA &
  $\approx30\%$ &
  6 GHz \\  
\cite{gao2017photon} &
  \begin{tabular}[c]{@{}c@{}} Photon-trapping Si PD \end{tabular} &
  \begin{tabular}[c]{@{}c@{}}  800-900 nm \end{tabular} & 
  5 V &
  60 pA &
  $52\%$$^l$ &
  19.5 GHz$^b$ \\  
\cite{chatterjee2019high} &
  \begin{tabular}[c]{@{}c@{}}SiN-on-SOI \\ evanescently coupled
  \end{tabular} &
  850 nm &
  \begin{tabular}[c]{@{}c@{}}20 V \\ 10 V
  \end{tabular} &
  \begin{tabular}[c]{@{}c@{}}75 nA \\ 46 nA
  \end{tabular} &
  \begin{tabular}[c]{@{}c@{}}$42\%$$^b$ \\ $29\%$$^b$
  \end{tabular} &
  \begin{tabular}[c]{@{}c@{}}14 GHz \\ 15.5 GHz
  \end{tabular} \\
\cite{pour2017high} &
  \begin{tabular}[c]{@{}c@{}}Grating-assisted \\ lateral PIN Si-PD on SOI\end{tabular} &
  850 nm &
  \begin{tabular}[c]{@{}c@{}}20 V$^c$ \\ 12 V$^d$
  \end{tabular} &
  \begin{tabular}[c]{@{}c@{}}1 nA$^c$ \\ 0.4 nA$^d$
  \end{tabular} &
  \begin{tabular}[c]{@{}c@{}}$47\%$$^{b,c}$ \\ $22\%$$^{b,d}$
  \end{tabular} &
  \begin{tabular}[c]{@{}c@{}}4.7 GHz$^c$ \\ 15 GHz$^d$
  \end{tabular} \\ 
\cite{bernard2021top} &
  \begin{tabular}[c]{@{}c@{}} SiON to Si substrate \end{tabular} &
  \begin{tabular}[c]{@{}c@{}}  850 nm \end{tabular} &  
  3 V &
  N.A. &
  $44.3\%$ &
  N.A. \\  
\cite{chen2018integration} &
  \begin{tabular}[c]{@{}c@{}} GaAs MSM PD \end{tabular} &
  \begin{tabular}[c]{@{}c@{}}  850 nm \end{tabular} &  
  2 V &
  22 pA &
  $17\%$$^b$ &
  20 GHz \\  
  \hline
\end{tabular}%
}
\begin{tablenotes}
\item$V_{RB}$: reverse bias, $I_{dark}$: dark current, EQE: external quantum efficiency, BW: 3-dB optoelectronic bandwidth
\item$^a$discrete red (640 nm), green (532 nm), blue (405, 445, and 488 nm) wavelengths
\item$^b$inferred from the paper
\item$^c$intrinsic width of 5 \textmu m
\item$^d$intrinsic width of 2 \textmu m
\item$^e$measured at $V_{RB}=5$ V
\item$^f$$\lambda=$400-560 nm
\item$^g$$\lambda=$560-600 nm
\item$^h$$\lambda=$600-748 nm
\item$^i$$\lambda=$748-955 nm
\item$^j$$\lambda=$ 405-532 nm
\item$^k$based on the light reaching Si junction
\item$^l$at $\lambda=$ 850 nm
\item N.A.: not available
     \end{tablenotes}
\end{table}
\clearpage

\begin{table}[!t]
\centering
\caption{Comparison of submicrometer-wavelength (VIS and NIR) integrated photodetectors operating with avalanche or photoconductive gain}
\label{tab:comparison_APD}
\resizebox{\textwidth}{!}{%
\begin{tabular}{ccccccccc}
\hline
\textbf{Ref.} &
  \textbf{Type} &
  \textbf{Wavelength} &
  \textbf{$V_{RB}$} &
  \textbf{$I_{dark}$} &
  \textbf{R} &
  \textbf{M} &
  \textbf{BW} &
  \textbf{GBP}
  \\ \hline
This work &
 \begin{tabular}[c]{@{}c@{}}SiN-on-SOI (VIS-PD) \end{tabular} & 
 532 nm &
 \begin{tabular}[c]{@{}c@{}}13.5 V\end{tabular} &
 \begin{tabular}[c]{@{}c@{}}$<$9 nA\end{tabular} &
 \begin{tabular}[c]{@{}c@{}} 0.30 A/W$^d$ \end{tabular}  &
\begin{tabular}[c]{@{}c@{}}$10^2-10^4$$^e$ \end{tabular} & 
 \begin{tabular}[c]{@{}c@{}}N.A\end{tabular} &
\begin{tabular}[c]{@{}c@{}}N.A.\end{tabular} \\ 
This work &
 \begin{tabular}[c]{@{}c@{}}SiN-on-SOI (NIR-PD) \end{tabular} & 
 785 nm &
 \begin{tabular}[c]{@{}c@{}}13.5 V\end{tabular} &
 \begin{tabular}[c]{@{}c@{}}$<$20 nA\end{tabular} &
 \begin{tabular}[c]{@{}c@{}}0.25 A/W$^d$\end{tabular} &
 \begin{tabular}[c]{@{}c@{}}25$^f$\\ 29$^g$ \end{tabular} &
 \begin{tabular}[c]{@{}c@{}} 11 GHz$^f$ \\ 13 GHz$^g$\end{tabular} &
 \begin{tabular}[c]{@{}c@{}} 278 GHz$^f$\\ 374 GHz$^g$ \end{tabular} \\
\cite{lin2022monolithically} &
 \begin{tabular}[c]{@{}c@{}}SiN-on-Si bulk \\ evanescently coupled \end{tabular} &
  405 nm &
  13 V$^a$ &
  $\approx1$ \textmu A$^a$ &
  0.20 A/W $^{a,d}$ &
  $46\pm14$ &
  $3.7\pm1.6$ GHz &
  $173\pm30$ GHz\\
\cite{yanikgonul2021integrated} &
  \begin{tabular}[c]{@{}c@{}} SiN end-coupled (lateral doping) \\ SiN end-coupled (interdigitated doping)\end{tabular} &
  685 nm &
 \begin{tabular}[c]{@{}c@{}}20 V \\ 18 V \end{tabular} &
 \begin{tabular}[c]{@{}c@{}}0.12 \textmu A \\ 0.31 \textmu A \end{tabular} &
 \begin{tabular}[c]{@{}c@{}}$0.83\pm0.05$ A/W \\ $0.63\pm0.01$ A/W \end{tabular} &
 \begin{tabular}[c]{@{}c@{}}12.3  \\ 2.25 \end{tabular} &
 \begin{tabular}[c]{@{}c@{}}19.1 GHz  \\ 14.8 GHz \end{tabular} &
 \begin{tabular}[c]{@{}c@{}}$234\pm25$ GHz  \\ $33\pm1$ GHz \end{tabular} \\
\cite{cuyvers2022heterogeneous} &
  \begin{tabular}[c]{@{}c@{}}Heterogeneously integrated \\Si PD \end{tabular} &
  \begin{tabular}[c]{@{}c@{}}  775 nm\end{tabular} &  
  45 V &
  0.4 \textmu A$^a$ &
  0.19 A/W$^d$ &
  $\approx10$$^a$ &
  6.8 GHz$^a$ &
  68 GHz\\  
\cite{pour2017high} &
  \begin{tabular}[c]{@{}c@{}}Grating-assisted \\ lateral PIN Si-PD on SOI\end{tabular} &
  850 nm &
  14 V &
  2 \textmu A &
  0.3 A/W &
  6 &
  16.4 GHz &
  98.4 GHz$^a$ \\
\cite{yanikgonul2022high} &
  \begin{tabular}[c]{@{}c@{}}SiN-on-SOI \\ evanescently coupled 
  \end{tabular} &
  850 nm &
  39 V &
  \begin{tabular}[c]{@{}c@{}}6 nA$^b$ \\ 21 nA$^c$ 
  \end{tabular} &
  \begin{tabular}[c]{@{}c@{}}0.32 A/W$^b$ \\ 0.71 A/W$^c$ 
  \end{tabular} &
  \begin{tabular}[c]{@{}c@{}}9.1$^b$ \\ 11.4$^c$ 
  \end{tabular} &
  \begin{tabular}[c]{@{}c@{}}12.8 GHz$^b$ \\ 7.7 GHz$^c$ 
  \end{tabular} &
  \begin{tabular}[c]{@{}c@{}}116.6 GHz$^b$ \\ 87.8 GHz$^c$ 
  \end{tabular} \\
\cite{chatterjee2020compact} &
  \begin{tabular}[c]{@{}c@{}}Ring-coupled Si MSM PD \\ SiN-on-SOI\end{tabular} &
  850 nm &
  5 V &
  13.1 nA &
  0.78 A/W & 
  1.3$^{a,h}$ &
  6.2 GHz$^{a,i}$ &
  8.06 GHz$^{i}$\\    
    \hline
\end{tabular}%
}
\begin{tablenotes}
\item$V_{RB}$: reverse bias, $I_{dark}$: dark current, R: responsivity, M: avalanche or photoconductive gain, BW: 3-dB optoelectronic bandwidth, GBP: gain-bandwidth product
\item$^a$inferred from the paper
\item$^b$$L_{PD}=100$ \textmu m
\item$^c$$L_{PD}=500$ \textmu m
\item$^d$responsivity in linear-mode 
\item$^e$depends on the optical power, Fig. \ref{fig:AvalancheGain_VIS} 
\item$^f$$L_{PD}=200$ \textmu m, $P_{in,PD}=$ 2.9 \textmu W
\item$^g$$L_{PD}=400$ \textmu m, $P_{in,PD}=$ 2.9 \textmu W
\item$^h$photoconductive gain
\item$^i$bandwidth was inferred from the reported full-width at half-maximum (FWHM) of the time response at 10 V, it was not available for 5 V
\item N.A.: not available
     \end{tablenotes}
\end{table} 

\begin{backmatter}
\bmsection{Acknowledgments} This work was supported by the Max Planck Society. The authors thank Hakan Deniz, Norbert Schammelt, and Stuart Parkin at the Max Planck Institute of Microstructure Physics for transmission electron microscope imaging. The authors also thank Salih Yanikgonul for helpful discussions. 
\bmsection{Disclosures} J.C.M. and J.K.S.P. are currently employed by Lightmatter, a company developing optical interconnects based on photonic integrated circuits. The other authors declare no conflicts of interest. 
\bmsection{Data availability} Data underlying the results presented in this paper are not publicly available at this time but may be obtained from the authors upon reasonable request. 
\end{backmatter}

\bibliography{Main_refs}

\end{document}